\documentclass[a4paper,11pt]{article}
\usepackage{geometry}
\usepackage[colorlinks=true,urlcolor=black,linkcolor=blue,citecolor=blue,bookmarks=false,breaklinks]{hyperref}
\usepackage{amsfonts,amsmath,amssymb}
\usepackage{subcaption}
\usepackage{graphicx}
\usepackage[font=small,labelfont=bf]{caption}
\usepackage{cite}
\usepackage[page,toc,titletoc,title]{appendix}
\usepackage{url}
\usepackage[OT1]{fontenc}
\usepackage{authblk}
\usepackage{fancyhdr}
\usepackage{booktabs}

\graphicspath{{figures/}}

\renewcommand*{\thepage}{\footnotesize\arabic{page}}
\providecommand{\keywords}[1]{\small{\textbf{Keywords:} #1}}

\newcommand{\indi}[1]{\varepsilon^{(#1)}(t)}
\newcommand{\fig}[1]{Fig.~\ref{#1}}

\title{\bf Correlations in Motion: A Simple Response-Based Analysis of Traffic Flow}

\author{Sebastian Gartzke\thanks{sebastian.gartzke@uni-due.de}, Shanshan Wang, Thomas Guhr and Michael Schreckenberg}
\affil{\textit{Faculty of Physics, University Duisburg--Essen, Lotharstra\ss e 1, 47048 Duisburg, Germany}}

\date{\today}
 
\begin{document}
\maketitle

\begin{abstract}
\noindent Why does a traffic jams form out of nowhere, and why does it stretch for kilometers even after the initial cause is passed? This study examines how congestion moves and spreads across motorways using a surprisingly simple method: response functions. These functions are based purely on data and show how changes in traffic flow, density, and velocity are connected over time and space. Using real-world data from German motorways, we track how traffic reacts to earlier disturbances, capturing the waves of slowing and accelerating that drivers experience in stop-and-go traffic. The results demonstrate how congestion propagates and how its rhythm can be measured and predicted. Unlike complex traffic models, this approach requires no simulations or assumptions about driver behavior. It works directly from the information provided by the road. The goal is clear: to understand congestion better so that we can manage it more effectively and perhaps spend less time stuck in it.

\vspace{\baselineskip} \noindent \keywords{response functions, traffic congestion, vehicular traffic, time series analysis, complex system}
\end{abstract}

\noindent\rule{\textwidth}{1pt}
\vspace*{-1cm}
{\setlength{\parskip}{0pt plus 1pt} \tableofcontents}
\noindent\rule{\textwidth}{1pt}

\frenchspacing
\clearpage
\section{Introduction}
\label{sec1}

Traffic congestion is a widespread feature of modern mobility that affects the daily routines of millions. Yet, its emergence and evolution remain difficult to fully predict and control. On motorways in particular, traffic behaves like a complex, many-body system where local disruptions can lead to widespread congestion. Despite advances in traffic science, understanding how disturbances travel across a network and how they affect key traffic variables, such as flow, velocity, and density, remains a major challenge.

Everyone who has ever been caught in a traffic jam has likely asked the same questions: Where is this coming from? Why does it spread? And why does it sometimes dissolve just as quickly as it formed? The experience of driving can be surprisingly disconnected from the bigger picture of what is actually happening along the road network. A slowdown in one area can spread through the system, affecting traffic kilometers away. However, understanding how and why this happens, and how one traffic variable influences another, is not trivial. Scientists have long known that variables such as velocity, vehicle density, and flow are closely linked. Yet, knowing that a link exists is not the same as being able to measure it clearly.

This raises a fundamental question: How can we quantify these interdependencies and understand how congestion propagates in time and space, without relying on overly complex models? Can we capture these dynamics using only data, and do so in a way that is robust, transparent, and computationally efficient? A method that meets these criteria could benefit not only researchers, but also planners and engineers, by providing practical tools for interpreting traffic data and identifying emerging patterns. It would also support the broader effort to move from descriptive observations to measurable and actionable findings in traffic systems research.

To address this challenge, we turn to response functions, a method that measures how changes in traffic observables at one location follow congestion events at another. Applied in financial market analysis, response functions quantify the impact of past events on a system's behavior across different time scales. In finance, response functions have been used to reveal the complexity of market reactions and the temporal structure of cross-asset dependencies. Notably, they have enabled significant advances in the understanding of cross-impact mechanisms~\cite{Wang2016a,Wang2017,Benzaquen2017}, liquidity effects~\cite{Wang2017,Henao2021}, sectoral dependencies~\cite{Wang2016b}, and the non-Markovian nature of price fluctuations~\cite{Philippe2003}. Previous trades and correlated sector responses influence future price movements in measurable and often nonlinear ways. These findings motivate the use of similar techniques in other complex systems, including transportation.

By conditioning responses on defined congestion indicators, we can extract how congestion affects downstream traffic, and do so in a way that reveals both propagation velocity and oscillation patterns, all from empirical time series data.
While many established traffic models, such as fluid dynamical theories~\cite{Lighthill55,Richards56}, gas kinetic approaches~\cite{Newell55,Prigogine71}, and car following models~\cite{Gazis61,Gipps81,Newell02}, have provided deep insights into the mechanics of traffic flow and have been instrumental in understanding the conditions under which congestion arises, they often rely on specific model assumptions and parameter calibrations that may limit their generalizability. Cellular automata~\cite{NaSch_1992,Schadschneider_1993,Schreckenberg_1995,Barlovic_1998} and three phase traffic theory~\cite{Kerner_2002,Kerner_2004,Kerner_2009} have successfully reproduced a wide range of emergent traffic patterns, including stop and go waves and capacity drops, often in high spatial resolution. However, these models can become complex when extended to real world networks or when applied to empirical validation.

Response functions offer a complementary perspective. Rather than aiming to model the full system from first principles, they empirically measure how traffic variables at one location respond to past disruptions elsewhere, directly from data. This allows for a transparent and model independent quantification of how congestion propagates in space and time. In this way, response functions do not replace established models but rather enhance our ability to test, interpret, and constrain them using observable system behavior. Especially for questions involving the temporal ordering and spatial spread of disturbances, response functions provide a robust, data driven framework that can be used in conjunction with more detailed simulation based approaches.

Recent studies have shown their potential in capturing temporal and spatial effects of congestion using velocity as the primary observable~\cite{Wang2020,Gartzke2022,Wang2023}. However, their broader capacity to quantify the interplay between flow, density, and velocity, and to systematically assess the spatiotemporal spread of congestion, remains largely unexplored. In this study, we address that by applying response functions to real motorway data from North Rhine–Westphalia, Germany, and investigate their ability to characterize traffic dynamics across multiple observables.

The paper is organized as follows. Sec.~\ref{sec2} presents the data set studied and introduces response functions and related indicators of congestion. Sec.~\ref{sec3} introduces the locations used for the study. The results are presented in Sec.~\ref{sec4}. Sec.~\ref{sec5} briefly summarizes and concludes our results.

\section{Data description and methods}
\label{sec2}
For our study we use traffic data collected by inductive loop detectors on motorways in North Rhine-Westphalia (NRW), Germany. In Sec.~\ref{sec21} we describe the data used for the study and present the methods applied in Sec.~\ref{sec21}.

\subsection{Data}
\label{sec21}

Inductive loop detectors deployed on motorways are designed to gather information about the traffic situation by recording essential observables on each lane \cite{Treiber2010}. These observables are the traffic flow rate~$q$ and the average velocity~$v$. The measurements are conducted at each section over time intervals of one minute, and the aggregated data are presented in units of vehicles/h and km/h, respectively. Furthermore, the detectors are capable of differentiating between cars and trucks. Our study is limited to workdays, (weekends and public holidays are excluded) resulting in a total of $N=243$ workdays from December 2016 to December 2017 included in our data set.

In this study the data from all lanes at each section are aggregated as a single effective lane. For each lane $l$ at each section $s$ a traffic density
\begin{equation}
	\rho_{sl}^{\text{(m)}}(t) =  \frac{q_{sl}^{\text{(m)}}(t)}{v_{sl}^{\text{(m)}}(t)}, 
	\label{eq21_1}
\end{equation}
is determined from the traffic flow $q_{sl}^{\text{(m)}}(t)$ and the velocity $v_{sl}^{\text{(m)}}(t)$ for cars and trucks with $\text{m} = \text{car},\text{tr}$. The sum of the respective quantities of cars and trucks provides the total traffic flow and density
\begin{align}
	q_{sl}(t) &= q_{sl}^{(\text{car})}(t) + q_{sl}^{(\text{tr})}(t), \label{eq21_2} \\ 	
	\rho_{sl}(t) &= \rho_{sl}^{(\text{car})}(t) + \rho_{sl}^{(\text{tr})}(t), 	\label{eq21_3}
\end{align}
for each lane. The total effective flow~$q_s(t)$ and density~$\rho_s(t)$ for each section are obtained by summing over all lanes. The corresponding effective velocity~$v_s(t)$ at each section is then given by
\begin{equation}
	v_s(t) = \frac{q_{s}(t)}{\rho_{s}(t)} = \frac{\sum_{l} q_{sl}(t)}{\sum_{l} \rho_{sl}(t)}.
	\label{eq21_4}
\end{equation}
Although the use of the hydrodynamic relation in Eq.~\eqref{eq21_1} is a standard procedure for determining traffic densities from loop detector data, it should be noted that this method is an approximation. Deriving a spatial quantity from a temporal quantity can potentially introduce systematic errors \cite{Treiber2010}. By employing the time averaged velocity in the calculation with Eq.~\eqref{eq21_1}, rather than the spatially averaged velocity, the actual density is typically slightly underestimated. Nevertheless, in the majority of cases, this approximation is deemed sufficient since the primary characteristics of macroscopic traffic effects are preserved.

\subsection{Response functions and congestion indicators}
\label{sec22}
This contribution focuses on the analysis of response functions
\begin{equation}
	\tilde{R}_{ij}^{(x)}(\tau) = \frac{\sum\limits_{t=1}^{T-\tau} (x_i(t+\tau) - x_i(t)) \varepsilon_j(t)}{\sum\limits_{t=1}^{T-\tau} \varepsilon_j(t)} = \frac{\left\langle \Delta x_i(t,\tau) \ \varepsilon_j(t) \right\rangle_t}{\left\langle \varepsilon_j(t) \right\rangle_t},
	\label{eq22_1}
\end{equation}
which yields the response, or the average change, of a traffic observable~$x$ at section~$i$ at time~$t+\tau$ conditioned on congestion indicated by~$\varepsilon_j(t)$ at section~$j$ at a previous time~$t$. In regard to the study by Wang et al. \cite{Wang2023}, we take the initial step of extending the analysis to encompass not only the responses of velocities ($x = v$), but also those of flow ($x=q$) and density ($x=\rho$). In a second step, the complex nature of freeway traffic is taken into account by varying the indicator~$\varepsilon_j(t)$. An alternative approach would be to define response functions in which the unconnected part $\left\langle \Delta x_i(t,\tau)\right\rangle_t \left\langle \varepsilon_j(t) \right\rangle_t$ is subtracted. Here we find it useful to use the definition in~\eqref{eq22_1}.

\begin{figure}[htb]
	\begin{center}
		\includegraphics[width=0.4\textwidth]{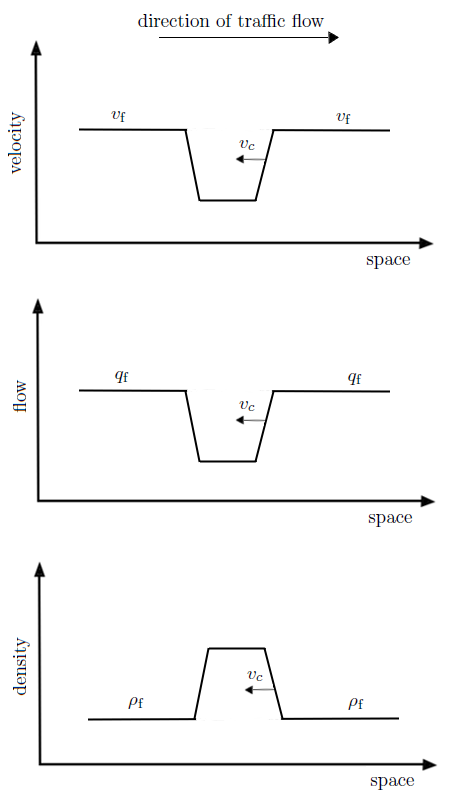}
		\caption{Schematic representation of congestion propagation against the direction of free traffic flow with velocity $v_\text{c}$ at a fixed time $t$. The homogeneous distribution of velocity~$v_\text{f}$, flow~$q_\text{f}$ and density~$\rho_\text{f}$ during a free flow phase is perturbed by a (density) shock wave. The sketch was drawn with Paint.net \cite{paintnet}.}
		\label{fig:wave_schematic}
	\end{center}
\end{figure}

On the one hand, in a relatively simple approach congestion can be described as a (density) shock wave propagating upstream through free flow traffic phases \cite{Lighthill55,Richards56}. In contrast, more recent traffic theory approaches demonstrate the existence of diverse phases (or states) of congestion and their transitions, which are more complex to describe, e.g. three-phase traffic theory \cite{Kerner_2004,Kerner_2009}. Considering the results of Wang et al. \cite{Wang2023}, we have developed an understanding of the characteristics that response functions may encompass in relation to congestion phases. Our analysis will reveal the capacity of responses to reflect the correlations between the fundamental traffic observables, namely velocity, flow, and density, during phases of congestion propagation, as illustrated in \fig{fig:wave_schematic}.
As compared to the response functions in Ref.~\cite{Wang2023}, we here refine the analysis by conditioning on velocity intervals. To this end, we analyse responses that incorporate the four different indicator functions
\begin{equation}
	\begin{aligned}
		& \varepsilon^{(0)}_j(t)  =
		\begin{cases}
			1, & \text{if } 0 \leq v_j(t) \leq 20 \ \text{km/h}, \\
			0, & \text{else,}
		\end{cases}, \ 
		& \varepsilon^{(1)}_j(t)  & =
		\begin{cases}
			1, & \text{if } 20 \ \text{km/h} < v_j(t) \leq 40 \ \text{km/h}, \\
			0, & \text{else,}
		\end{cases} 
		\\ & \varepsilon^{(2)}_j(t) =
		\begin{cases}
			1, & \text{if } 40 \ \text{km/h} < v_j(t) \leq 60 \ \text{km/h}, \\
			0, & \text{else,}
		\end{cases}, \
		& \varepsilon^{(3)}_j(t) & =
		\begin{cases}
			1, & \text{if } 0 \leq v_j(t) \leq 60 \ \text{km/h}, \\
			0, & \text{else.}
		\end{cases}
	\label{eq22_2}
	\end{aligned}
\end{equation}
Fine-tuned indicator functions enable the capture of different phases of congestion occurrence, such as stop-and-go traffic, wide moving jams, and others. In Eq.~\eqref{eq22_2}, $\indi{0}$ indicates heavily congested roadways with low velocity range. Indicators $\indi{1}$ and $\indi{2}$ are designed to indicate situations such as stop-and-go traffic or phases between two congestion waves. The introduction of the velocity ranges of the previous indicators results in $\indi{3}$, which represents a more inclusive threshold for the indication of congestion for comparison.

We focus on the analysis of traffic responses to congestion during rush hours. Specifically, we examine either morning rush hours~(mr) from 06:00 to 11:00 or afternoon rush hours~(ar) from 14:00 to 19:00, depending on the individual cases to be examined. The time lag in Eq.~\eqref{eq22_1} is constrained to a maximum of $\tau = 300 \ \text{min}$, covering the entirety of the rush hour interval. To avoid random events like accidents or construction sites affecting the results, we average response functions over the specified work days in this study. We proceed as follows: For each case under examination, an indicator section $j$ is selected, and all given time series of velocities $v_j(t)$ are transformed into four time series of indicators $\varepsilon_j^{(y)}(t)$, with $y=0,1,2,3$, according to Eq.~\eqref{eq22_2}. For each day and each pair of sections~$ij$ the corresponding response functions~$\tilde{R}_{ij}^{(x)}(\tau)$ are calculated for all observables (velocity, flow and density) for all four types of indicators. In each case, the calculated responses are averaged over all workdays, yielding the averaged response function $R_{ij}^{(x)}(\tau)$, with $x=v,q, \rho$ respectively, for the given indicator.

\section{Studied locations}
\label{sec3}
Interchanges are important nodes in transportation networks and typically serve more than one function. They provide a connection between two (or more) motorways, allow travelers to change direction, and in some cases serve as an entry point to the motorways network. They also have drawbacks during heavy traffic periods. Due to their structure, combined with the fact that on- and off-ramps act as bottlenecks \cite{Kerner_2004,Kerner_2009}, motorway interchanges are very likely to be a source of traffic congestion during rush hours. This is particularly true for interchanges that connect highly trafficked motorways. In addition to inducing congestion, interchanges have the potential to spread congestion to other parts of the motorway network. Congestion can spread through on-ramps and off-ramps to connected motorways or urban traffic networks, resulting in a deterioration of the current traffic situation.
\begin{figure}[!htbp]
	\centering
	\begin{tabular}[t]{cc}
		\begin{tabular}{c}
			\smallskip
			\begin{subfigure}[c]{0.48\textwidth}
				\centering
				\includegraphics[width=1\textwidth]{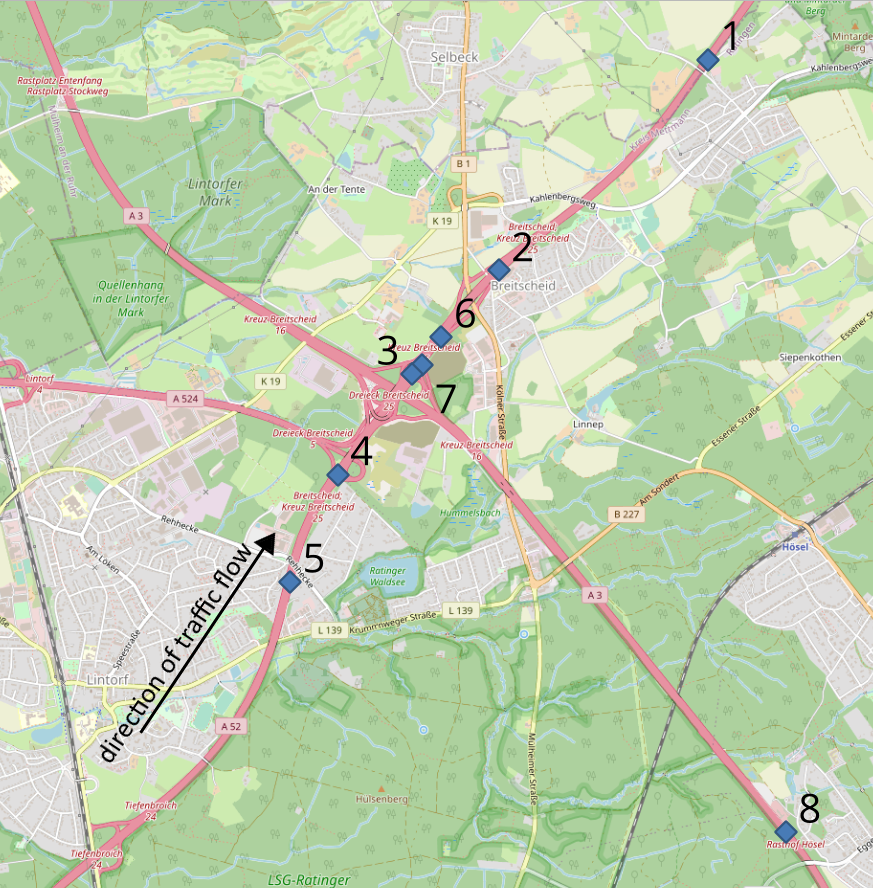}
				\caption{Breitscheid}
				\vspace{\baselineskip}
				\label{fig:map_breitscheid}
			\end{subfigure}\\
			\begin{subfigure}[c]{0.48\textwidth}
				\centering
				\includegraphics[width=1\textwidth]{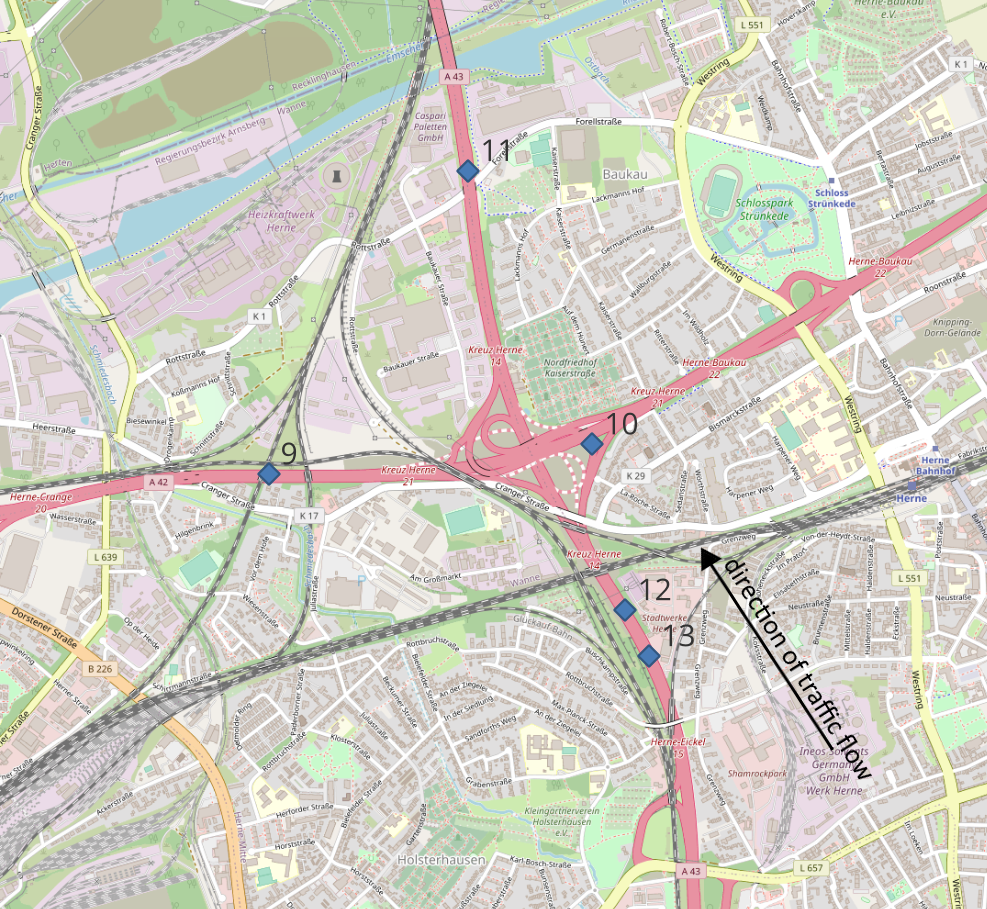}
				\caption{Herne}
				\label{fig:map_herne}
			\end{subfigure}
		\end{tabular}
		&
		\begin{subfigure}[c]{0.48\textwidth}
			\centering
			\smallskip
			\includegraphics[width=1\linewidth,height=1.5\textwidth]{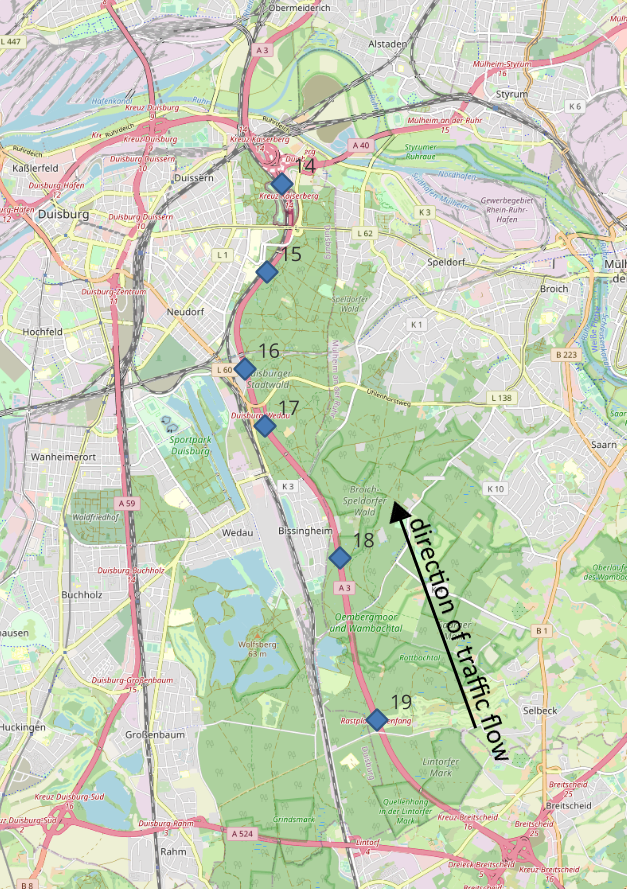}
			\caption{Motorway A3 between Breitscheid and Kaiserberg} 
			\label{fig:map_kaiserberg}
		\end{subfigure} \\
	\end{tabular}
	\caption{Locations of loop detectors (blue markers) for all three study cases. (a) Motorway interchange Breitscheid, (b) motorway interchange Herne and (c) motorway A3 between interchange Breitscheid and Kaiserberg. The numbering of sections is chosen in such a way that the direction of traffic flow goes from lager section numbers to smaller section numbers. All map data and map tiles are provided by OpenStreetMap \textcopyright OpenStreetMap Contributors and are licensed under ODbL v1.0 by the OpenStreetMap Foundation \cite{OSML,OpDbLi}.}
	\label{fig:map_locations}
\end{figure}

\begin{figure}[!htbp]
	\centering
	\begin{subfigure}[b]{1\textwidth}
		\centering
		\includegraphics[width=\textwidth]{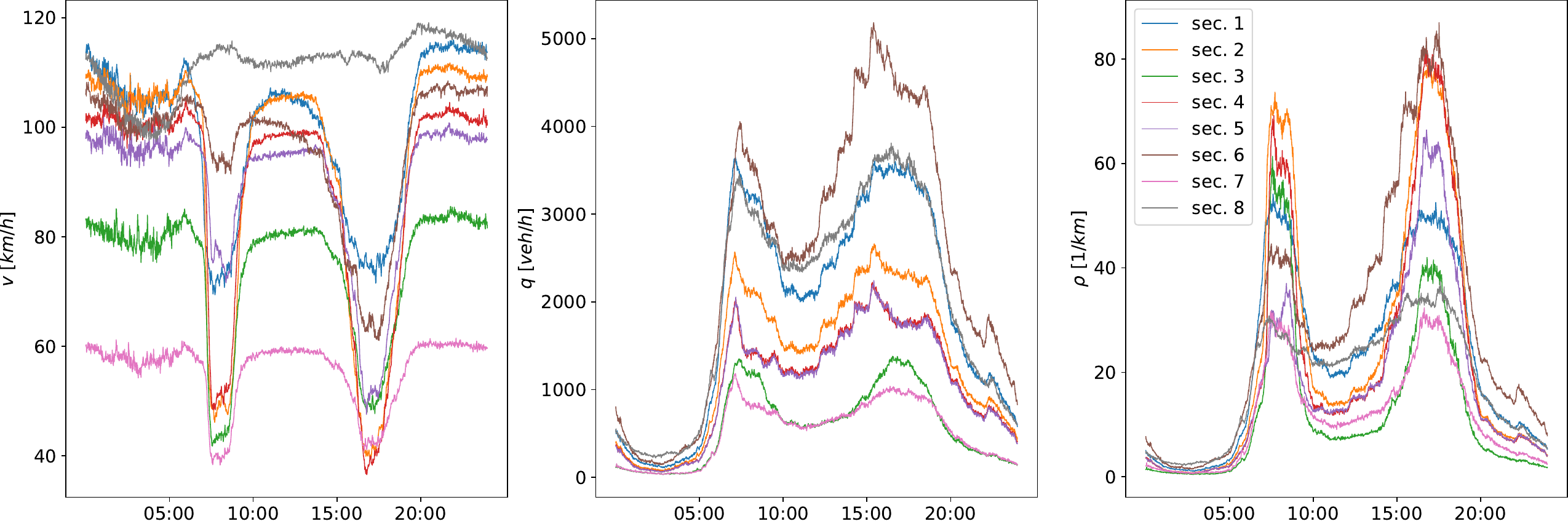}
		\caption{Breitscheid}
		\vspace{\baselineskip}
		\label{fig:avg_data_breitscheid}
	\end{subfigure}
	
	\begin{subfigure}[b]{1\textwidth}
		\centering
		\includegraphics[width=\textwidth]{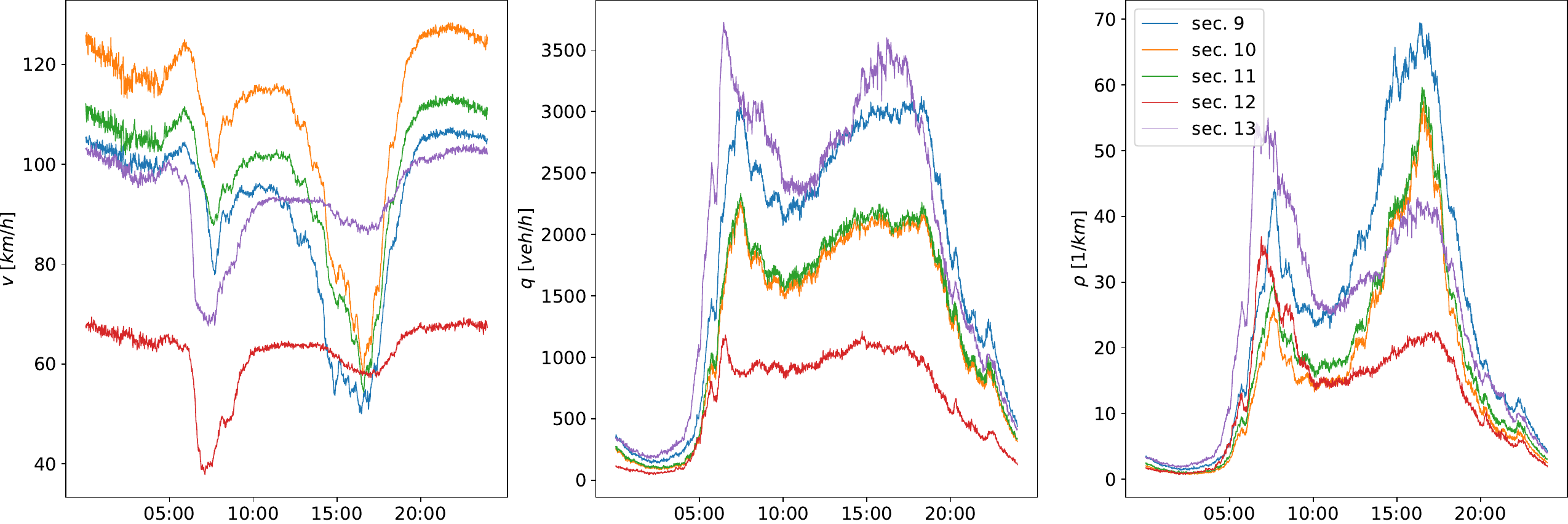}
		\caption{Herne}
		\vspace{\baselineskip}
		\label{fig:avg_data_herne}
	\end{subfigure}
	
	\begin{subfigure}[b]{1\textwidth}
		\centering
		\includegraphics[width=\textwidth]{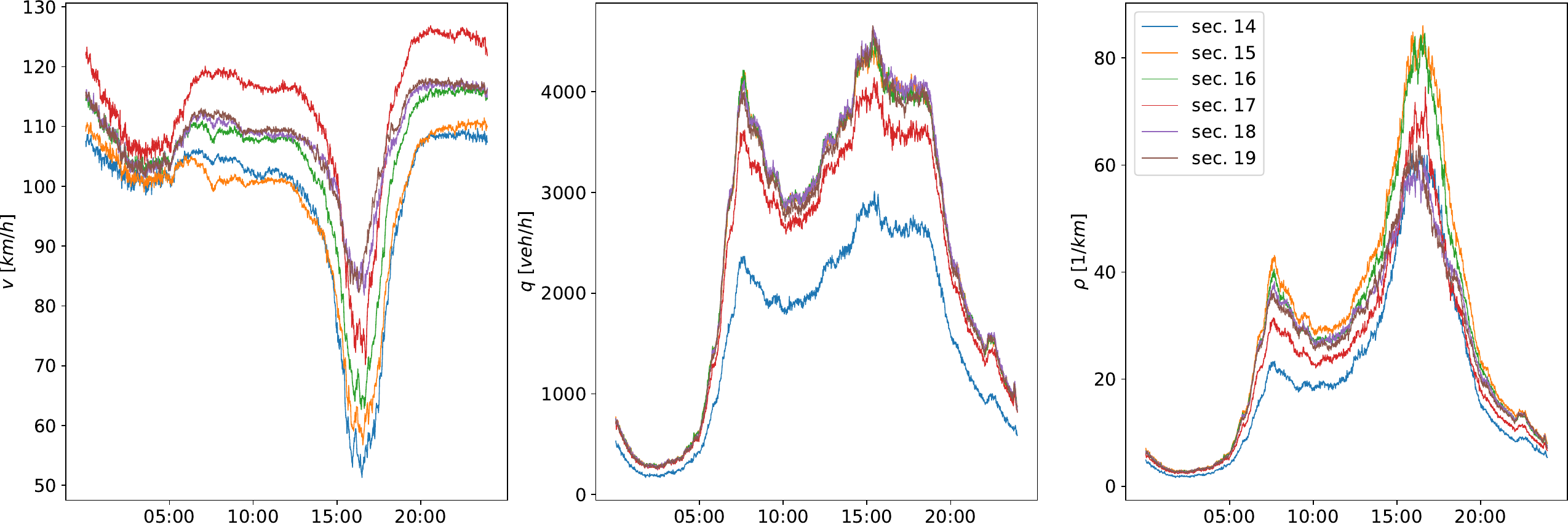}
		\caption{Motorway A3}
		\label{fig:avg_data_kaiserberg}
	\end{subfigure}
	\caption{Time evolution of effective flow $q$, velocity $v$ and density $\rho$ at all sections averaged over all $N=243$ workdays. The averaged data yields a representation for the usual traffic situation on workdays at the three different motorway segments shown in Fig.~\ref{fig:map_locations}. Distinct breakdowns in the averaged velocity time-series during the rush hours clearly indicate the occurrence of congestion on regular basis.}
	\label{fig:avg_data_locations}
\end{figure}

With regard to the characteristics mentioned above, we consider motorway interchanges and connected motorways for our study. We explore three different, heavily trafficked motorway segments in NRW, Germany. Two of these segments are motorway interchanges while the third is a part of motorway A3 connecting two interchanges. All three studied locations including the sections with loop detectors are shown in Fig.~\ref{fig:map_locations}. The numbering of sections is chosen in a way that the direction of traffic flow goes from larger section numbers to smaller section numbers. As a consequence, congestion propagates from smaller section numbers to larger section numbers. The calculation of responses is done on spatially subsequent sections. In addition, \fig{fig:avg_data_locations} illustrates the overall traffic situation on workdays with averaged data for all studied locations. Here we would like to take the opportunity to point out that the velocity in Figs. 3 and 4 in \cite{Gartzke2022} are given in the unit of veh/min instead of veh/h.

Figure~\ref{fig:map_locations}(a) shows the Breitscheid interchange, which connects three motorways, namely A52, A3 and A524. The A3 is a longer-range motorway that connects large cities such as Cologne (in the south), Duisburg and Oberhausen (in the north). Motorway A52 is a shorter-range motorway that connects the cities of Düsseldorf (in southwest direction) and Essen (in northeast direction). As shown in Fig.~\ref{fig:avg_data_locations}(a), the segment of the A52 in northeast direction, going from Düsseldorf to Essen, is heavily trafficked during rush hours and congestion is present on a daily basis. The traffic situation is complicated by the fact that traffic participants can exit and enter the motorway network before passing section 6 and section~2, respectively. 
Section~6 is a special case due to the general structure of the Breitscheid interchange. It is located on a separate lane which routes the traffic flow from the connected motorways and merges it with the traffic on the main lanes of A52 between section~2 and section~1 (see supplementary map in \fig{fig:Breitscheid_zoom} in Appendix~\ref{ap_map_breitscheid}). Section~1 will be the indicator section for the investigations performed.

The Herne interchange, shown in Fig.~\ref{fig:map_locations}(b), connects motorway A42 and A43. While the former connects large cities of the Ruhrgebiet (Ruhr area), such as Gelsenkirchen, Bottrop, Oberhausen and Duisburg (from east to west), the latter connects the city of Herne with Wuppertal, Bochum, Recklinghausen and Münster (from south to north). The total amount of traffic within the Herne intergange is not quite as high as in the case of the Breitscheid interchange, but Fig.~\ref{fig:avg_data_locations}(b) clearly demonstrates the occurrence of congestion on a regular bases during rush hours. Depending on the case under study, sections 11 and 9 will be the indicator sections for our investigation.

The third location is the segment of motorway A3 between the Breitscheid and Kaiserberg interchanges, see Fig.~\ref{fig:map_locations}(c). This part of the A3 is heavily trafficked northbound during afternoon rush hours on a daily basis. At the Kaiserberg interchange , near section 14 and (spatially) right next to an off-ramp, the number of main lanes is reduced from three to two. This reduction acts as an additional bottleneck, reducing the capacity of the motorway and causing heavy congestion on a regular basis. Figure~\ref{fig:avg_data_locations}(c) demonstrates how severe the congestion is on average and how far the traffic breakdown propagates to the south. Due to the traffic situation within the Kaiserberg interchange, we choose section 14 as the indicator section for our investigation.

\section{Results}
\label{sec4}
\subsection{A comparison of responses of velocity, flow and density}
\label{sec41}
\begin{figure}[!htbp]
	\centering
	\begin{subfigure}[b]{1\textwidth}
		\centering
		\includegraphics[width=0.8\textwidth]{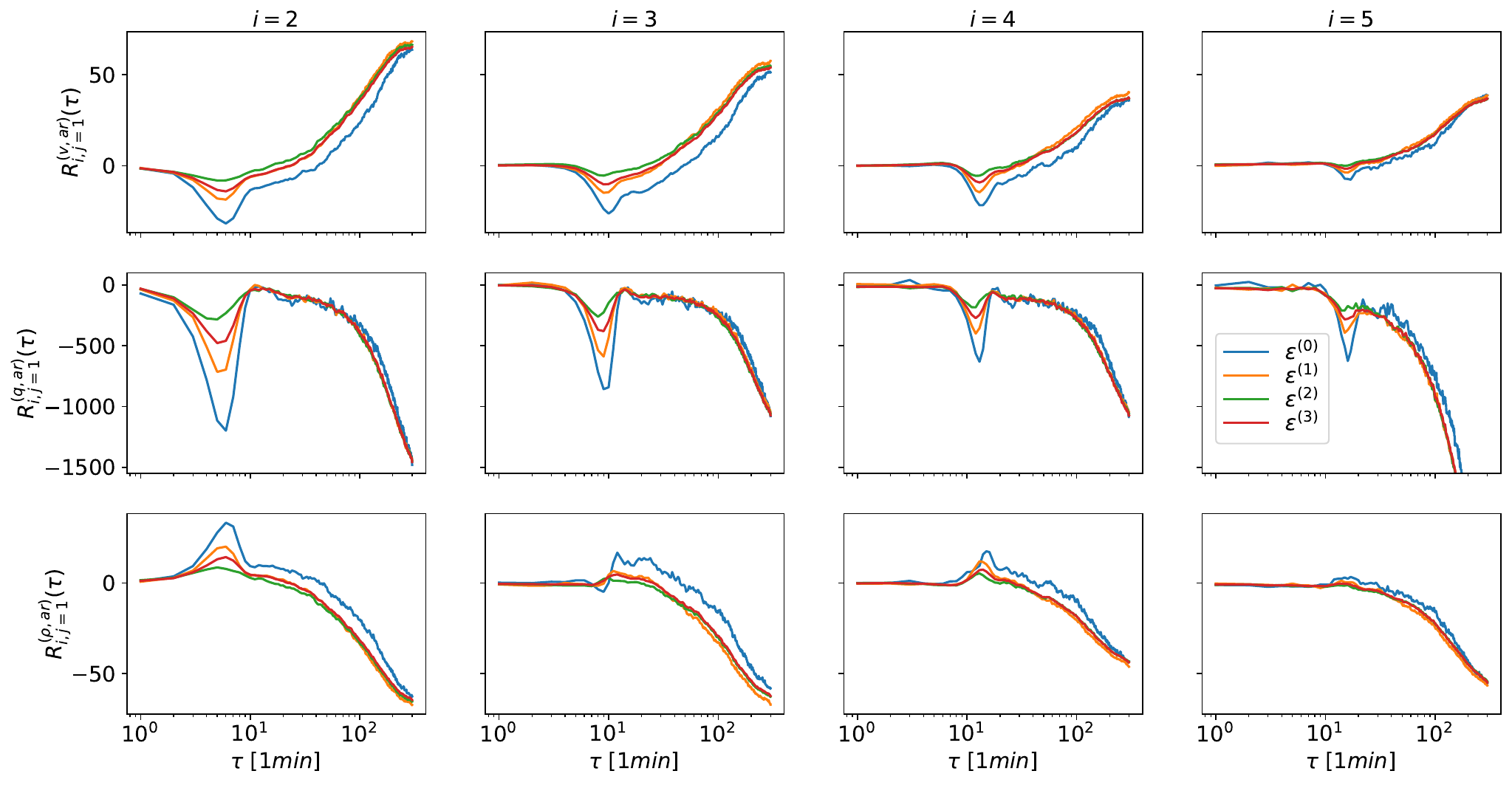}
		\caption{Breitscheid}
		\vspace{\baselineskip}
		\label{fig:generic_shape_A52}
	\end{subfigure}
	
	\begin{subfigure}[b]{1\textwidth}
		\centering
		\includegraphics[width=0.8\textwidth]{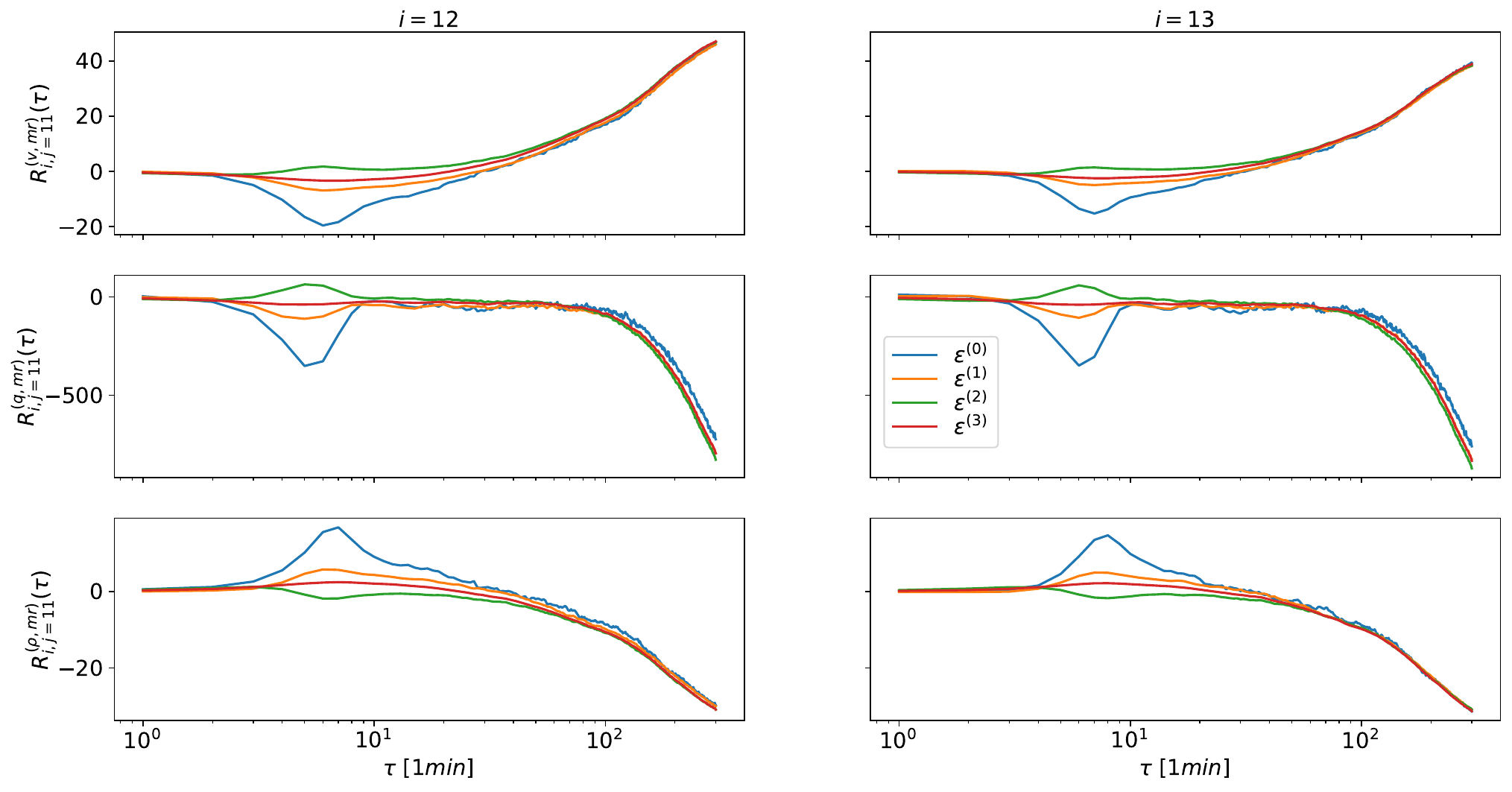}
		\caption{Herne}
		\vspace{\baselineskip}
		\label{fig:generic_shape_A43}
	\end{subfigure}
	
	\begin{subfigure}[b]{1\textwidth}
		\centering
		\includegraphics[width=0.8\textwidth]{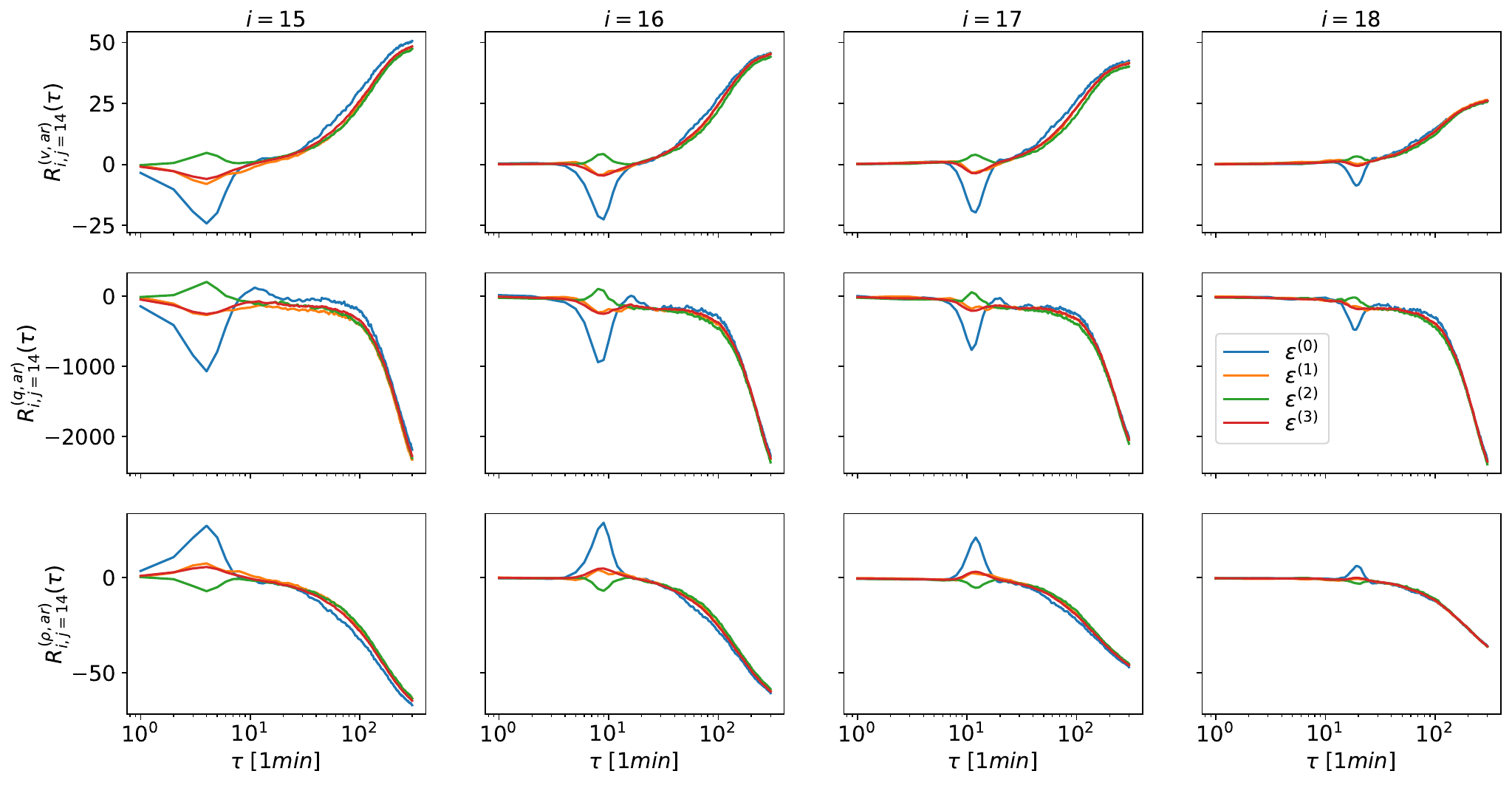}
		\caption{Motorway A3}
		\label{fig:generic_shape_A3}
	\end{subfigure}
	\caption{Velocity, flow and density response of spatially subsequent sections~$i$ to congestion at (a) section~$j=1$ at interchange Breitscheid during afternoon rush hours, (b) section $j=11$ at interchange Herne during morning rush hours and (c) section~$j=14$ on motorway A3 during afternoon rush hours.}
	\label{fig:generic_shape}
\end{figure}

It is useful to briefly relate our study to previous results from Wang et al. \cite{Wang2023}, in which interchange Breitscheid was already investigated. Figure~\ref{fig:generic_shape}(a) depicts the velocity, flow, and density responses of spatially subsequent sections to congestion at section 1 for all employed indicators during afternoon rush hours. Despite a different velocity range for indicator~$\indi{0}$ in the case of heavy congestion (up to 20 km/h instead of 10 km/h used in Ref.~\cite{Wang2023}), it is possible to identify similar behavior of velocity responses by comparison. The responses of neighboring sections~2~to~5 exhibit negative values, manifested as local minima or dips, for shorter time lags and rise to positive values, displaying continuous growth for longer time lags~$\tau$. The occurrence, amplitude, and width of response minimas depend on the distance between indicator section~$j$ and the responding section~$i$. As the distance increases, the minima occur at longer time lags with shrinking width and amplitude. This response behavior captures and reflects the velocity breakdown and, consequently, the upstream propagation of congestion. In comparison, the velocity responses for~$\indi{0}$ at interchange Breitscheid are found to be identical to those observed at interchange Herne in \fig{fig:generic_shape}(b) and on the A3 in \fig{fig:generic_shape}(c). This evidence suggests that the velocity responses for~$\indi{0}$ are independent of the specific location under study. We view this as an important identification of universalities in responses.

Prior to examining the responses of other indicators, we will initially investigate the flow and density responses to~$\indi{0}$ in \fig{fig:generic_shape}. Despite the inherently fluctuating nature of the flow on shorter time scales (a couple of minutes), the average responses exhibit a clear trend, including the occurrence of sharp local minima at larger time lags with increasing distance to the indicator section. This is observed in all three locations studied in \fig{fig:generic_shape}. The amplitudes of the negative responses of the nearest neighboring sections to the indicator section at Breitscheid ($i=2$) and motorway A3 ($i=15$) in \fig{fig:generic_shape}(a)~and~(c) corroborate the severity of the congestion occurring at these locations. Values between -1000~veh/h and -1200~veh/h clearly demonstrate the heavy traffic breakdown indicated by the average data in \fig{fig:avg_data_locations}. A comparison of velocity and flow responses with the density responses in \fig{fig:generic_shape} reveals that the expected courses have been met. As traffic congestion causes an increase in density at the local level, the respective density responses exhibit local maxima, occurring at larger time lags~$\tau$ for increasing distances from the indicator section. Furthermore, the scaling of the amplitudes in the density response reflects the severity of the traffic congestion at Breitscheid and on the A3, as illustrated in \fig{fig:generic_shape}(a)~and~(c). With values exceeding~20~veh/km for the nearest responding section, the local increase in density represents approximately a quarter of the average value shown in \fig{fig:avg_data_locations}. In summary, the local minima and maxima of all response types occur at the same time lag $\tau$ for the individual responding sections. Consequently, the responses of all three observables for~$\indi{0}$ capture the propagation of congestion and the correlations between velocity, flow, and density, as illustrated in \fig{fig:wave_schematic}.

A more detailed examination of the responses for the remaining congestion indicator reveals that the aforementioned correlations are also present in all three observables, with indicator~$\indi{2}$ representing a special case. Before discussing the latter, we will begin with the discussion of indicator~$\indi{3}$, which demonstrates the importance of choosing smaller velocity ranges to indicate congestion when studying traffic dynamics with response functions. While the minima and maxima of responses to~$\indi{3}$ are visible in case of interchange Breitscheid and motorway A3 in \fig{fig:generic_shape}(a)~and~(c), the response at interchange Herne does not exhibit any clear maxima or minima in \fig{fig:generic_shape}(b). Thus, it would not be possible to obtain any information about the traffic situation at interchange Herne. In addition, the amplitudes of the local minima and maxima of the responses to~$\indi{3}$ show either much smaller values, or values in range of indicator~$\indi{2}$ in case of motorway A3 and interchange Breitscheid. These findings demonstarte that the application of multiple indicators enables the extraction of more detailed information about the system. When the comparison of the remaining three indicators is restricted to the case of interchange Breitscheid, it can be observed that the amplitudes of the minima (maxima) decrease (increase) with increasing values of the velocity ranges. In terms of traffic flow theory, this behavior is expected. During the formation of a dense congestion phase, i.e. small distances between vehicles moving at low average velocity (e.g. $v < 20$ km/h ), an abrupt deceleration to low velocities induces a large drop in flow rate and an increase in local density. In contrast, less dense congestion allows vehicles to move at higher velocities, resulting in a smaller drop of flow rate and a decrease in local density. Under the assumption of vehicle number preservation, the hydrodynamic relation in Eq.~\eqref{eq21_1} takes this into account.

The response to indicator~$\indi{2}$ contrasts with the other responses at smaller~$\tau$ in the case of interchange Herne and motorway A3, as illustrated in \fig{fig:generic_shape}(b)~and~(c). Opposed to the expected behavior, the responses to indicator~$\indi{2}$ do not reach local minima for velocity and flow, nor do they reach local maxima for density. Instead, the responses exhibit local maxima for velocity and flow, and local minima for density. This is due to the presence of positive increments~$\Delta x_i(t,\tau)$ in Eq.~\eqref{eq22_1}. It represents vehicles accelerating during the congestion phase, which increases flow and reduces local density. Consequently, the correlations between the three observables remain, but appear in an opposite manner. However, the results depicted in \fig{fig:generic_shape} indicate a difference between the phases of congestion occurring at interchange Breitscheid and those occurring at interchange Herne and on motorway A3. Besides the acceleration phase captured by~$\indi{2}$, we find a similar case in the flow response for~$\indi{0}$ at interchange Breitscheid in \fig{fig:generic_shape}(c). A few minutes after the minimum, the response to~$\indi{0}$ propagates into a local maximum, indicating that vehicles are accelerating out of the traffic jam. Unfortunately, this only manifests in the flow response but not in the velocity or density responses to~$\indi{0}$. Although there are no distinct local maxima, there is an increasing trend in both responses. Overall, response functions are capable of capturing acceleration phases during congestion, which is an important finding.

\subsection{On- and off-ramp propagation of congestion and generic response features}
\label{sec42}

\begin{figure}[!htbp]
	\centering
	\begin{subfigure}[b]{1\textwidth}
		\centering
		\includegraphics[width=\textwidth]{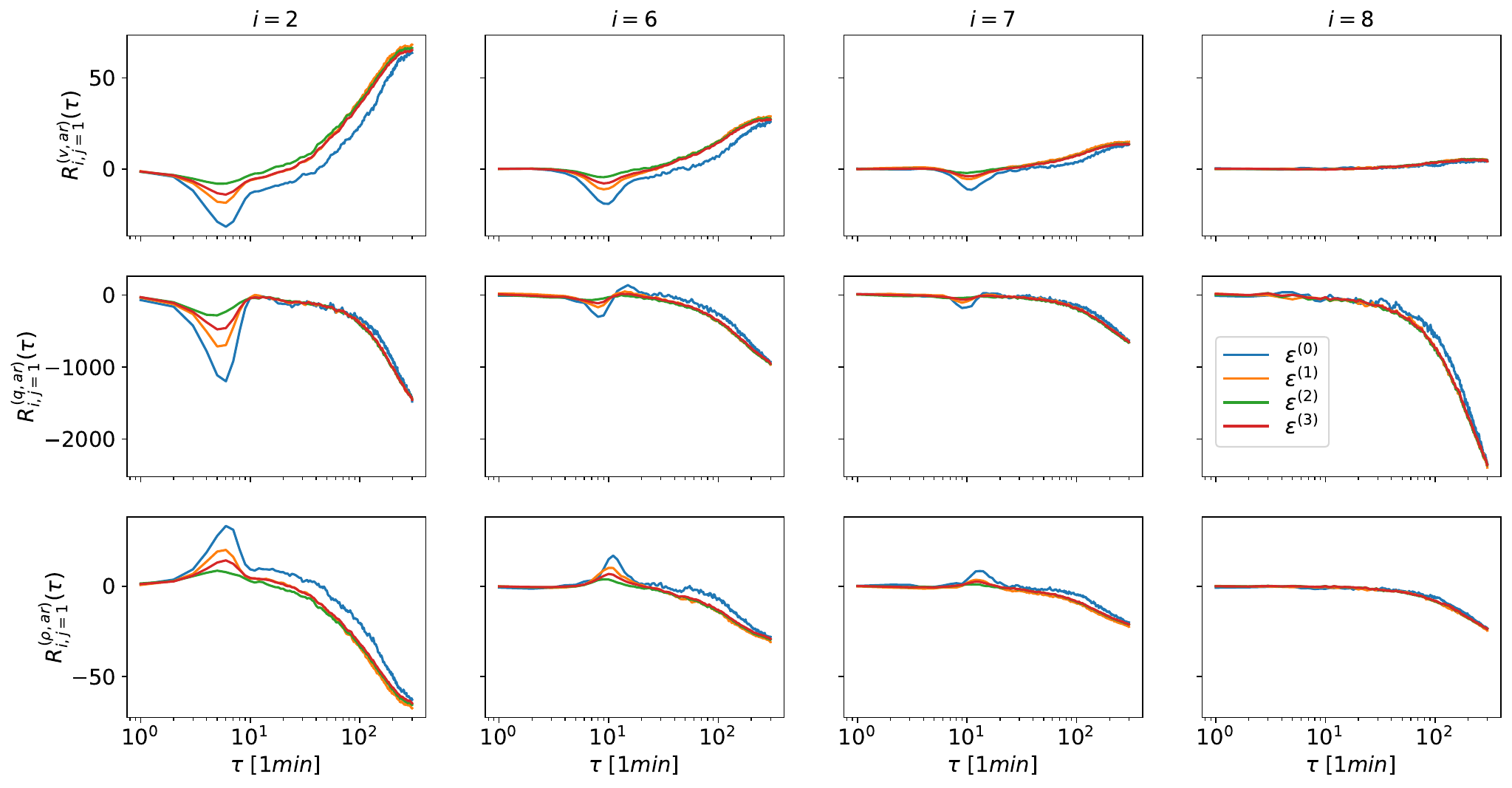}
		\caption{Breitscheid}
		\vspace{\baselineskip}
		\label{fig:direction_change_Breitscheid}
	\end{subfigure}
	
	\begin{subfigure}[b]{1\textwidth}
		\centering
		\includegraphics[width=\textwidth]{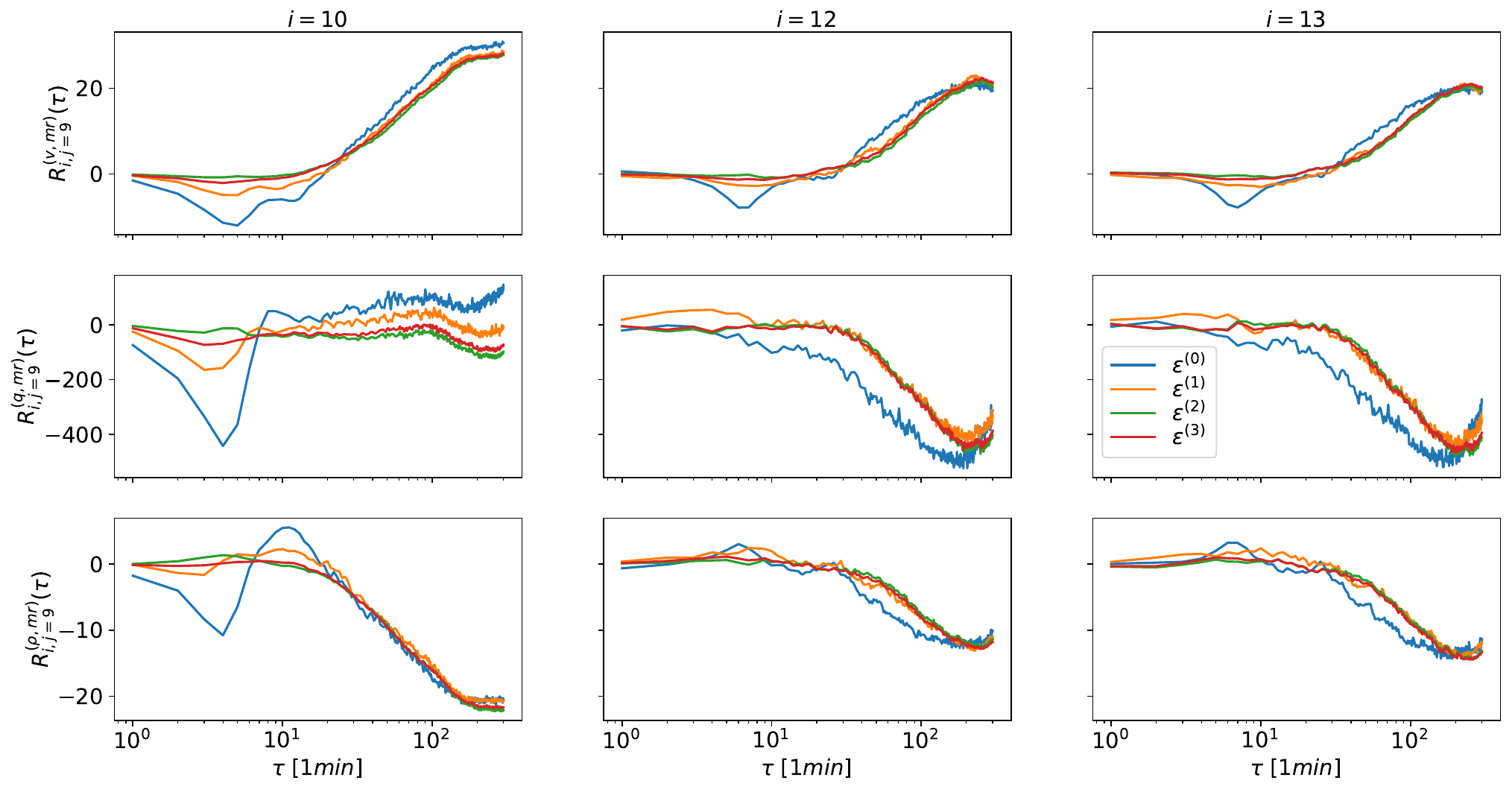}
		\caption{Herne}
		\vspace{\baselineskip}
		\label{fig:direction_change_Herne}
	\end{subfigure}
	
	\caption{Velocity, flow and density response of spatially subsequent sections $i$ to congestion at (a) section $j=1$ at interchange Breitscheid during afternoon rush hours and (b) section $j=9$ at interchange Herne during morning rush hours.}
	\label{fig:direction_change}
\end{figure}

Besides the results of Sec.~\ref{sec41} we are interested in the capability of response functions to detect the propagation of congestion through on- and off-ramps into other parts of the motorway network. Therefore we take a look at two cases at interchanges Breitscheid and Herne. In both cases we calculate the responses of spatially subsequent sections along an on-ramp or off-ramp, respectively, to an indicator section on a congested motorway. For interchange Breitscheid we chose section 1 as the indicator section and sections~2 and 6-8 as responding sections, where section~7 is located on the on-ramp leading traffic from motorway~A3 onto motorway~A52 and section~8 located on motorway~A3 (see \fig{fig:map_locations}(a)). In case of interchange Herne we select section~9 as the indicator section and sections~10, 12 and 13 as responding sections, where section 12~is located on the on-ramp leading traffic from motorway~A43 onto A42 (see \fig{fig:map_locations}(b)). Figure~\ref{fig:direction_change} contains the resulting responses for both cases.

As expected, all types of responses are identical for section 2 at interchange Breitscheid in
\fig{fig:direction_change}(a) (compared with \fig{fig:generic_shape}(a)). Furthermore, congestion can be seen moving through the on- or off-ramp, as seen in the responses of all three traffic variables at sections~6~and~7. In this case, the indicating minima (maxima) for $\indi{0}$ are the most distinct. The responses for section~8 do not show any local minima or maxima, indicating that the congestion does not propagate to section~8. It is also noteworthy that there is no available detector between sections~7~and~8.

In the case of the Herne interchange, a distinct indication of congestion can be observed propagating through the on- or off-ramp at section~12, as borne out by the velocity responses to indicator~$\indi{0}$ in \fig{fig:direction_change}(b). A comparison between the latter and the corresponding flow responses demonstrates that the flow data is unable to identify the propagation of congestion in this case. The flow responses of sections~12~and~13 exhibit a highly fluctuating course without any local extrema. In contrast, the responses of section~10 exhibit local minima (and maxima in the case of indicator~$\indi{2}$). An anomaly is observed when considering the corresponding density responses. The density responses of sections~12~and~13 to indicator~$\indi{0}$ exhibit local maxima that align with the courses of the corresponding velocity responses, indicating the propagation of congestion. In contrast, the density response at section~10 exhibits a local minimum, contrary to the expected local maximum.

Notwithstanding the possibility to clearly identify congestion propagation through on- or off-ramps in both cases in \fig{fig:direction_change}, we find the unexpected result of the density response exhibiting a maximum for section~10 at interchange Herne. Lacking any obvious traffic-related reason or systematic error, the unexpected result may be caused by missing values in the analyzed data. For the indicator section~9 about 14\% of that data are missing. For section~10 only~4\% of the flow data and~9\% of the velocity data are missing. The fluctuating courses of the flow responses also suggest a possible impact of low data quality on the anomalous results. Unfortunately, we cannot corroborate this point due to a lack of additional data, e.g. on construction sites or detector maintenance. At this point, it is clear that data quality and detector availability limit our study. Both the characteristics of the minima (and maxima) and the fluctuations in the response functions depend directly on the data quality. Due to occasional detector failures or temporary deactivations during construction, fully consistent data sets are difficult to obtain in practice. To mitigate this, we rely on averaging over a large number of independent days, which helps reduce the impact of isolated missing values and ensures stable response patterns. In addition, a higher resolution of the data could provide more accurate results, but the measurements of the detectors take place over a time interval of one minute. For this reason, the detectors themselves also limit our study.

Apart from the congestion propagation, the responses in \fig{fig:direction_change} and \fig{fig:generic_shape} show similar courses for large time lags $\tau$. Comparing all the cases studied, we find that the velocity responses increase while the flow and density responses decrease with increasing time lag $\tau$. Our results suggest that this behavior is a generic feature for the individual response types induced by the indicators defined in Eq.~\eqref{eq22_2}. These generic courses can be explained by the usual traffic situation represented by the averaged data in \fig{fig:avg_data_locations}. With increasing time, the traffic situation during rush hours gradually calms down, i.e. the number of traffic participants decreases. As a consequence, a decrease in density and flow allows higher velocities for the remaining traffic participants until the end of the rush hours. The generic feature we found is suitable for all responses in our study.

\subsection{Differences in congestion patterns}
\label{sec43}

In Sec.~\ref{sec41} we observe a deviation in the responses to indicator~$\indi{2}$ in comparison to the other indicators. The aforementioned deviant behavior clearly indicates an acceleration phase of the vehicles to moderate velocities for sections at interchange Herne and motorway A3 (in \fig{fig:generic_shape}~(b)~and~(c)), but not in the case of interchange Breitscheid (in \fig{fig:generic_shape}~(a)). Why is this so? From the perspective of traffic dynamics, the results indicate the presence of different congestion patterns. In the case of the A3 motorway or the interchange at Herne, vehicles are able to accelerate to higher velocities within the range of indicator~$\indi{2}$ before decelerating again due to the congestion. Consequently, larger gaps between traffic participants must be present, otherwise vehicles would be unable to accelerate to a higher velocity range. Therefore, the simultaneous occurrence of the extrema in the responses to indicators~$\indi{0}$~and~~$\indi{2}$ suggests that the congestion pattern at these locations is characterized by a stop-and-go nature, exhibiting acceleration-deceleration cycles. In contrast, the responses to indicators~$\indi{1}$~and~$\indi{2}$~at interchange Breitscheid do not exhibit any indication of an acceleration phase, i.e. local maxima, during the critical congestion period. This finding suggests that the congestion pattern is more dense and synchronized, with vehicles moving at low velocities (in the range of~$\indi{0}$~and~$\indi{1}$) while maintaining small distances to vehicles ahead of them.
\begin{figure}[!htbp]
	\centering
	\begin{subfigure}[b]{1\textwidth}
		\centering
		\includegraphics[width=\textwidth]{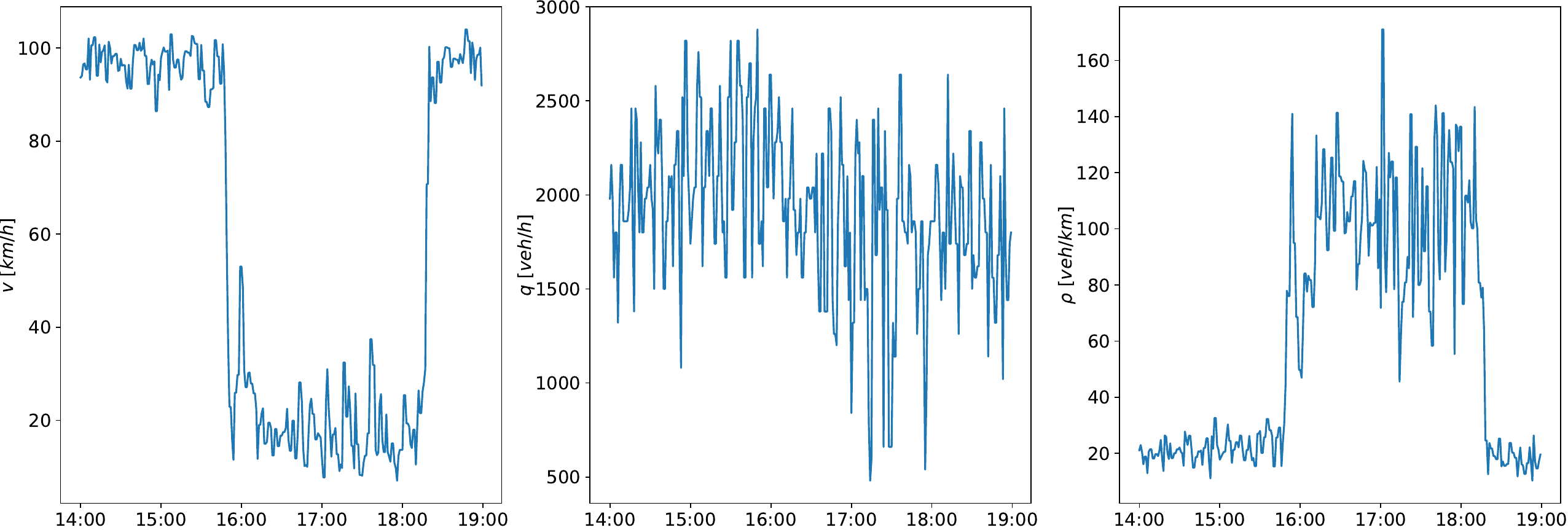}
		\caption{Breitscheid}
		\vspace{\baselineskip}
		\label{fig:data_example_A52}
	\end{subfigure}
	
	\begin{subfigure}[b]{1\textwidth}
		\centering
		\includegraphics[width=\textwidth]{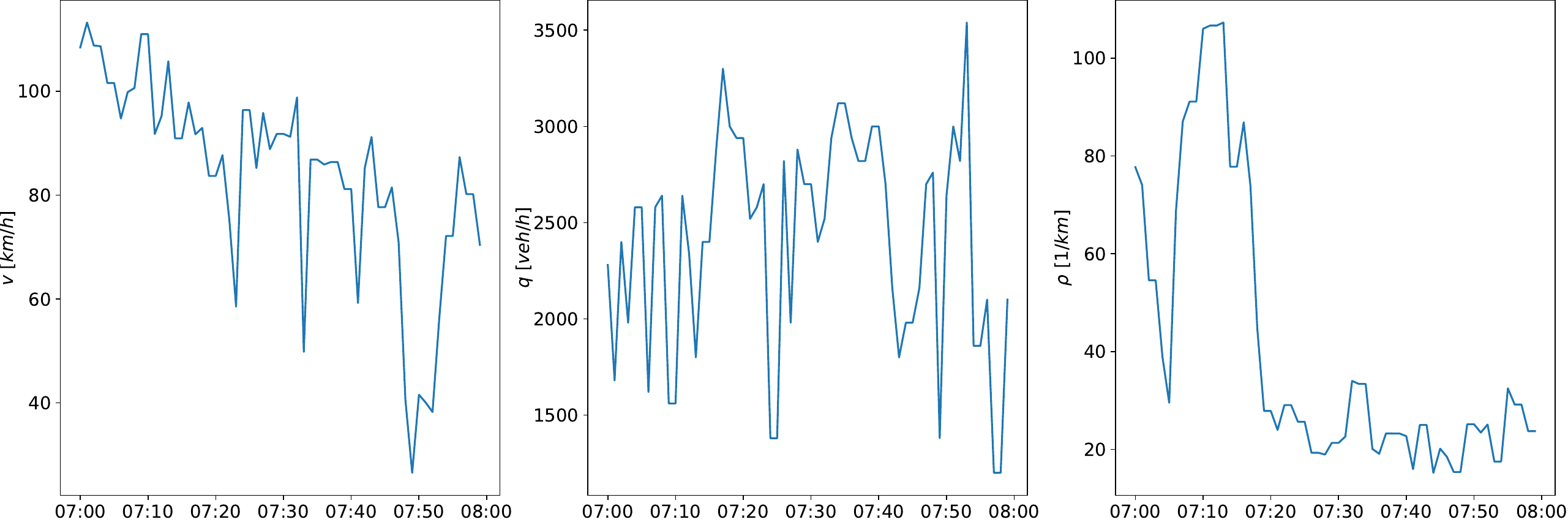}
		\caption{Herne}
		\vspace{\baselineskip}
		\label{fig:data_example_A43}
	\end{subfigure}
	
	\begin{subfigure}[b]{1\textwidth}
		\centering
		\includegraphics[width=\textwidth]{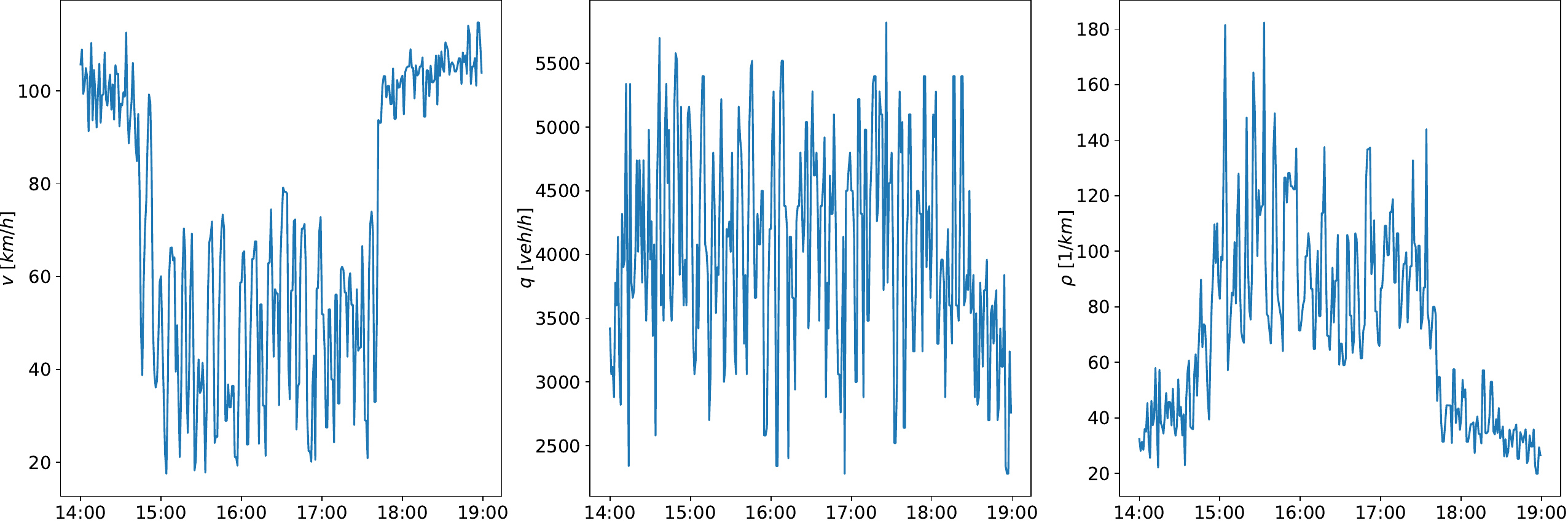}
		\caption{Motorway A3}
		\label{fig:data_example_A3}
	\end{subfigure}
	\caption{Sample of the daily data for (a) section~3 at interchange Breitscheid during afternoon rush hours, (b) section~12 at interchange Herne during morning rush hours and (c) section~15 on motorway A3 during afternoon rush hours. The data in (a) and (c) where recorded on Tuesday, December 13, 2016, the data in (b) on Wednesday, April 26, 2017.}  
\label{fig:data_example}
\end{figure}
This does not necessarily rule out a stop-and-go pattern of congestion, but the effect is not strong enough to be captured by the responses.

Given the nature of the data at our disposal, it is not possible to verify the aforementioned assumptions about the congestion pattern with information regarding the distances between traffic participants. Consequently, we consider the raw data. Figure~\ref{fig:data_example} illustrates a sample of the daily data, according to Eq.~\eqref{eq21_4}, for all locations under consideration. In all cases, the time period of the traffic breakdown can be discerned by examining the trends in velocity and density. Due to its fluctuating nature, the impact on traffic flow is not as readily apparent as with other observables. What the average responses already indicate is revealed by the velocity data (first column in \fig{fig:data_example}). While the velocity at section~3 at interchange Breitscheid is nearly always below the range of indicator~$\indi{2}$, the velocities at sections~12~and~15 at interchange Herne and on motorway A3, respectively, frequently reach (and exceed) the range of indicator~$\indi{2}$. Although a traffic oscillation, or stop-and-go pattern, is evident in the velocities at all three locations, the corresponding responses in \fig{fig:generic_shape} only show an acceleration phase, or local maxima, for indicator~$\indi{2}$ in the case of interchange Herne and motorway A3. Considering the velocity data for section~3 at Breitscheid (in \fig{fig:data_example}), one is led to anticipate a local maximum in the response to indicator~$\indi{1}$ due to the fact that the velocity frequently exceeds the range of indicator~$\indi{0}$. However, this is not the case.

\begin{figure}[!htbp]
\centering
\begin{subfigure}[b]{1\textwidth}
	\centering
	\includegraphics[width=\textwidth]{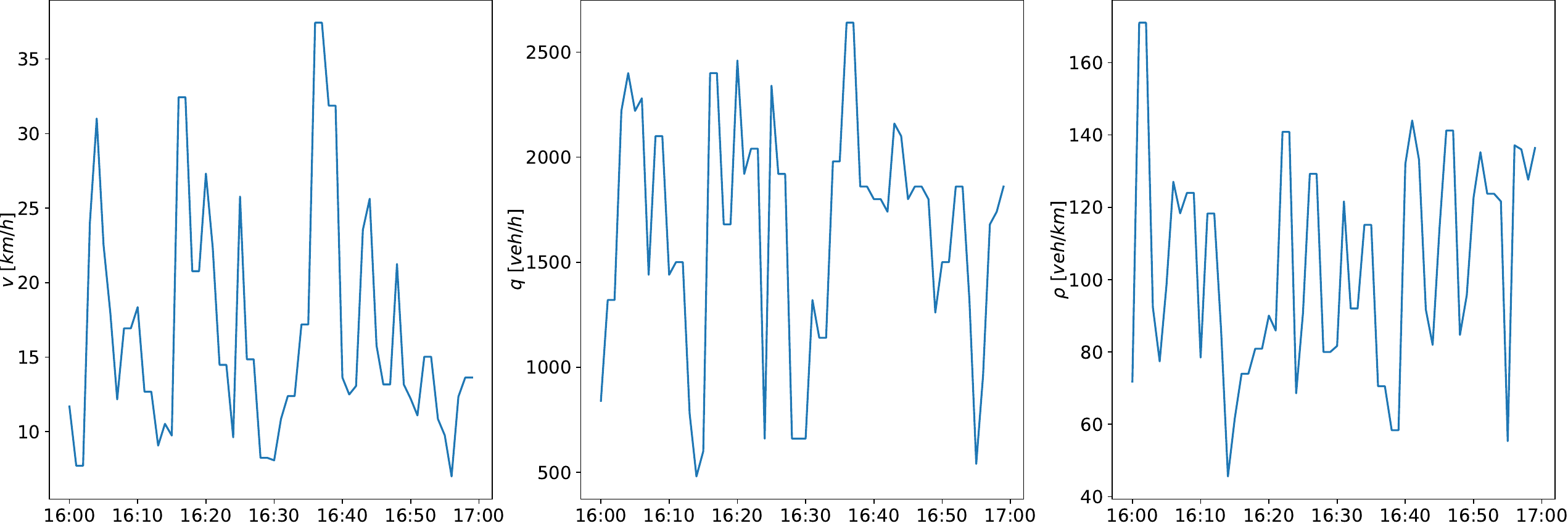}
	\caption{Breitscheid}
	\vspace{\baselineskip}
	\label{fig:data_example_A52_zoom}
\end{subfigure}

\begin{subfigure}[b]{1\textwidth}
	\centering
	\includegraphics[width=\textwidth]{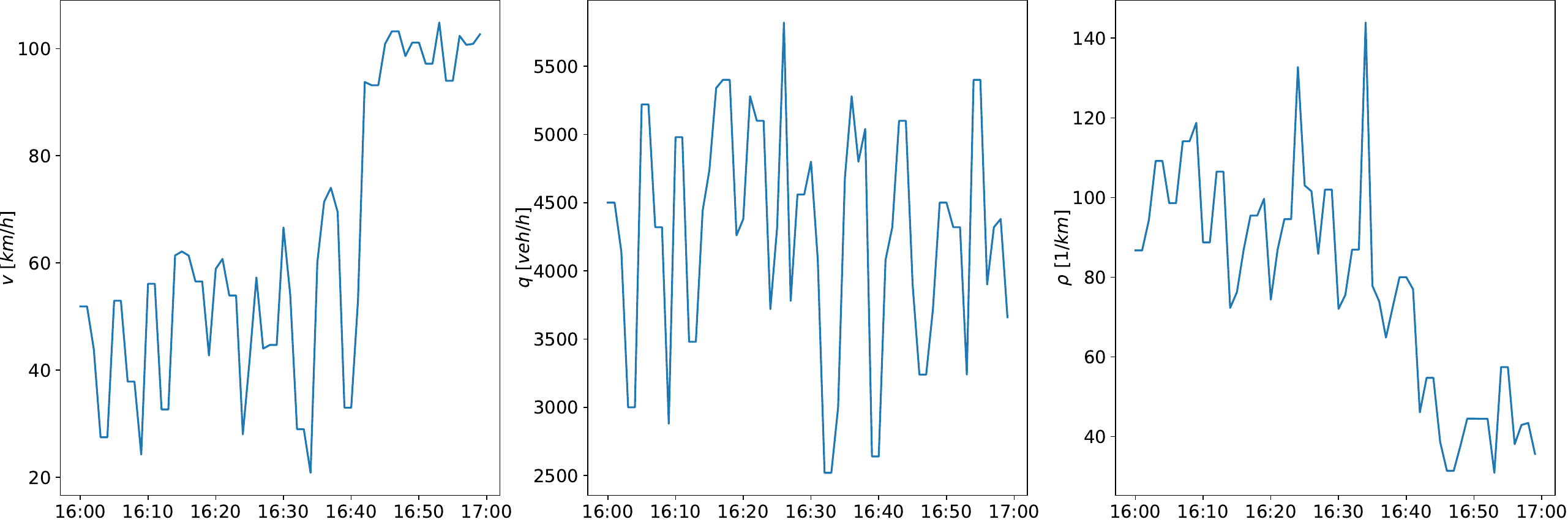}
	\caption{Motorway A3}
	\label{fig:data_example_A3_zoom}
\end{subfigure}
\caption{Zoom into the time period from 16:00 to 17:00 of the daily data from \fig{fig:data_example} for (a) section 3 at interchange Breitscheid and (b) section 15 on motorway A3.}
\label{fig:data_example_zoom}
\end{figure}

One possible explanation for the absence of a visible acceleration phase in the response to indicator~$\indi{1}$ in Breitscheid is the frequency with which the sample velocity distribution occurs, i.e. the latter is on average a rare case and the velocity does not exceed the range of indicator~$\indi{1}$ as often as in the example in \fig{fig:data_example}(a). Another possible explanation is a difference in the frequency of the traffic oscillation itself. Therefore, we take a closer look at the data sample in \fig{fig:data_example_zoom}. By enumerating the number of traffic oscillations, or the transition from one velocity minimum to the next (which, as illustrated in \fig{fig:data_example_zoom}, is approximately 10 km/h for Breitscheid and approximately 25 km/h for the A3), it is observed that there are five oscillations for the A3 and three oscillations for the Breitscheid interchange during the initial 30 minutes of the rush hour. Therefore, the temporal frequency of the traffic oscillations on the A3 is approximately twice that observed at the Breitscheid interchange.

Furthermore, a possible correlation may be established between the full width of the minima in the course of the responses and the periodicity of the traffic oscillation, in addition to the aforementioned statements. Given that the valleys or minima (hills or maxima) represent a sequence of deceleration and acceleration (acceleration and deceleration), it may be expected that their width (in relation to the $\tau$-axis) corresponds to the average time taken to traverse such a sequence during congestion. To test this assumption, we again consider the data sample shown in \fig{fig:data_example_zoom} and compare the period length of the traffic oscillation with the width of the valleys in the course of the responses. Accordingly, the range of velocities depicted in the samples is taken into account. With regard to the interchange at Breitscheid, a period of approximately ten minutes is observed for the traffic oscillation, with velocities ranging from 5 to 35 km/h, as illustrated in \fig{fig:data_example_zoom}(a). A comparison of this with the corresponding indicators~$\indi{0}$~and~$\indi{1}$ in \fig{fig:generic_shape}(a) reveals that the width of the valleys is also approximately 10 minutes, thus confirming the period. With the exception of the density response, for which the signal curve is blurred, the aforementioned coincidence is also observed in the flow. With regard to the A3 motorway, the sample depicted in \fig{fig:data_example_zoom}(b) predominantly exhibits velocities between 25 and 60 km/h, with an oscillation period of approximately five minutes. A comparison of the latter with the width of the valleys in the responses to indicators~$\indi{0}$~and~$\indi{1}$ in \fig{fig:generic_shape}(c) also demonstrates the coincidence in the case of velocity, flow, and density. These findings substantiate the aforementioned results for the temporal frequency of the traffic oscillations.

In conclusion, we briefly comment on the classification of the identified congestion patterns according to the three-phase traffic theory. A comprehensive examination of the identification of traffic phases or phase transitions is beyond the scope of this study. However, it is worthwhile to hint at the potential offered by response functions for this purpose. The aforementioned findings indicate that the congestion observed at interchange Breitscheid and on the A3 motorway is likely to be a wide-moving jam. This conclusion can be traced back to two fundamental properties of wide moving jams \cite{Kerner_2004}. On the one hand, a wide moving jam propagates upstream through any traffic state and bottleneck with a constant mean velocity of its downstream jam front. On the other hand, wide moving jams exhibit a notable longitudinal expansion. In the case of the interchange at Breitscheid, the responses indicate congestion with a spatial expansion of approximately $4.2$ km (the distance between sections~1~and~5), which propagates through a bottleneck at section~6 (see Sec.~\ref{sec42} and \fig{fig:Breitscheid_zoom} in Appendix~\ref{ap_map_breitscheid}). The responses to sections on the A3 motorway indicate that congestion propagates over an even longer distance, with a spatial extension of approximately $8.8$ km (between sections~14~and~19). These findings follow from the calculated responses without additional data, providing compelling evidence for the classification of the congestion as a wide moving jam.

With regard to the differentiation of congestion patterns, our study is not only limited by the quality of the data. Unfortunately the relationship between the widths of the local minima and the traffic oscillation cannot be established using the averaged data in \fig{fig:avg_data_locations}. Furthermore, the relationship is clearer for detectors close to the reference detector than for detectors that are further away. The width of the minima increases slightly for more distant detectors. Therefore, our approach is locally limited.
 
\subsection{The propagation velocity of congestion}
\label{sec44}
\begin{figure}[!htbp]
	\begin{center}
		\includegraphics[width=1\textwidth]{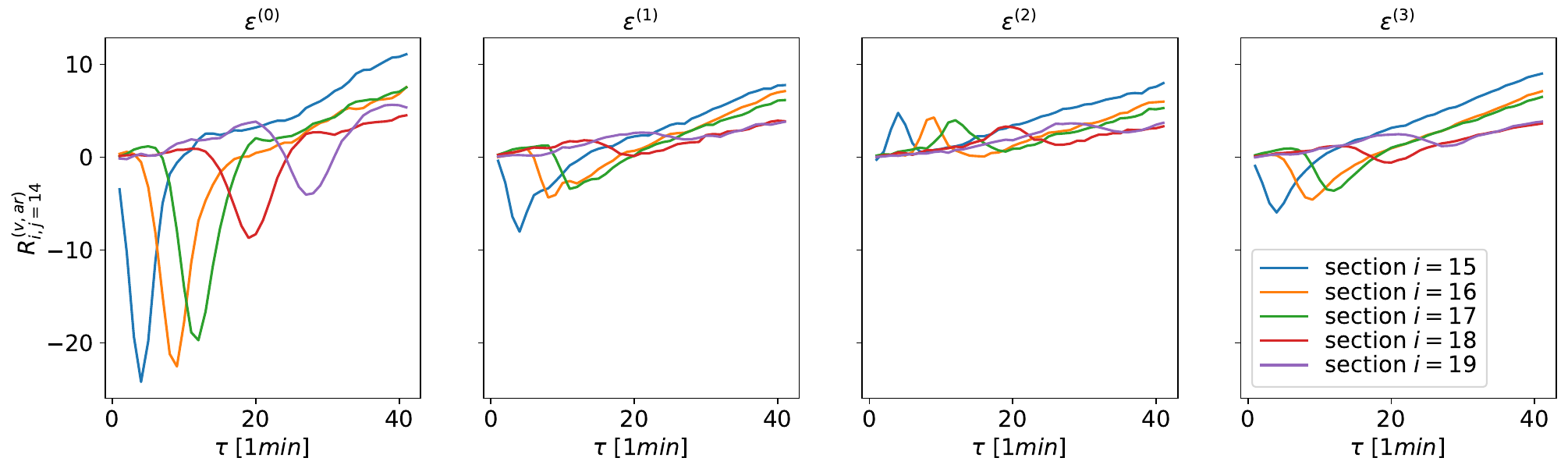}
		\caption{Velocity response of spatially subsequent sections $i$ to congestion at section $j=14$ on motorway A3 during afternoon rush hours broken down by the individual indicators $\varepsilon(t)$.}
		\label{fig:generic_shape_min_sample}
	\end{center}
\end{figure}

Figure~\ref{fig:generic_shape_min_sample} depicts the velocity responses for sections on motorway~A3, as previously illustrated in \fig{fig:generic_shape}(c), broken down by the individual indicators. This representation serves to emphasise the propagation nature of the minima and maxima, and therefore the propagation of congestion, in a particularly clear manner. This striking representation prompts the question of whether the propagation velocity~$v_c$ of congestion can be derived from our findings, which typically ranges between 15~km/h and 20~km/h~\cite{Kerner_2004,Treiber2010}. To answer the question, we approach to determine the propagation velocity of congestion from all types of responses for the Breitscheid interchange and the A3 motorway. A complete overview of all responses for both locations in the representation of \fig{fig:generic_shape_min_sample} is given in \fig{fig:generic_shape_min} in the Appendix~\ref{ap_generic_shape_min}. To determine the propagation velocities, we work out the positions of the corresponding local minima (or maxima) in the response curves for two sections $k$ and $l$ to obtain the time lag difference $\Delta\tau^{(\sigma)}_{kl}$. Dividing the distance $d_{kl}$ between the two sections by the corresponding time lag difference determines the propagation velocity $v_{\text{prop}}^{(x)}$ of the congestion:
\begin{equation}
	v_{\text{prop}}^{(x)} = \frac{d_{kl}}{\Delta\tau^{(\sigma,x)}_{kl}},
	\label{eq44_1}
\end{equation}
where $x=v,q,\rho$ denotes the type of response and $\sigma=\text{min,max}$ the type of local extremum that has to be taken into account.

\begin{figure}[!htbp]
	\centering
	\begin{subfigure}[b]{1\textwidth}
		\centering
		\includegraphics[width=\textwidth]{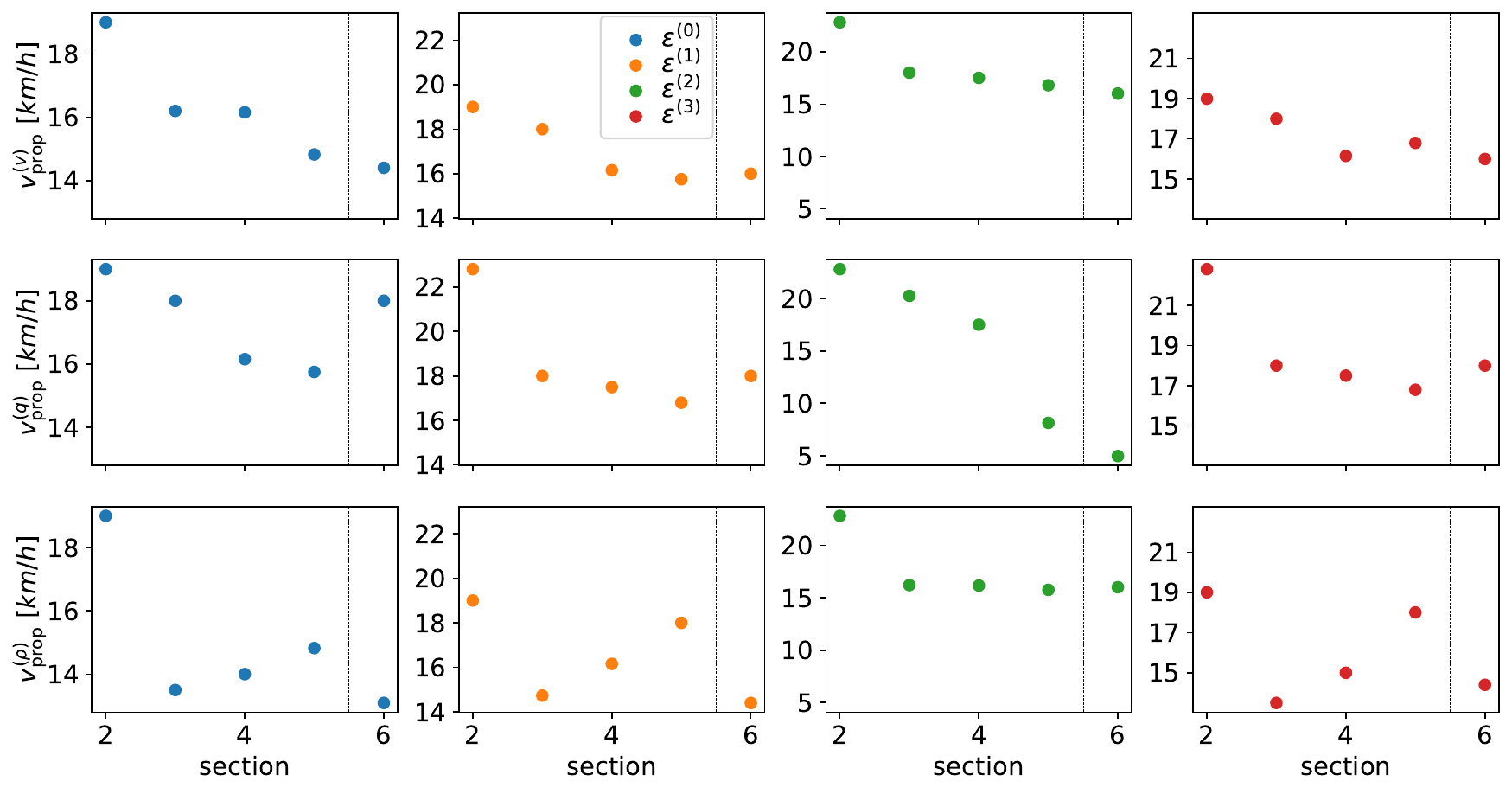}
		\caption{Breitscheid}
		\vspace{\baselineskip}
		\label{fig:prop_velo_A52}
	\end{subfigure}
	
	\begin{subfigure}[b]{1\textwidth}
		\centering
		\includegraphics[width=\textwidth]{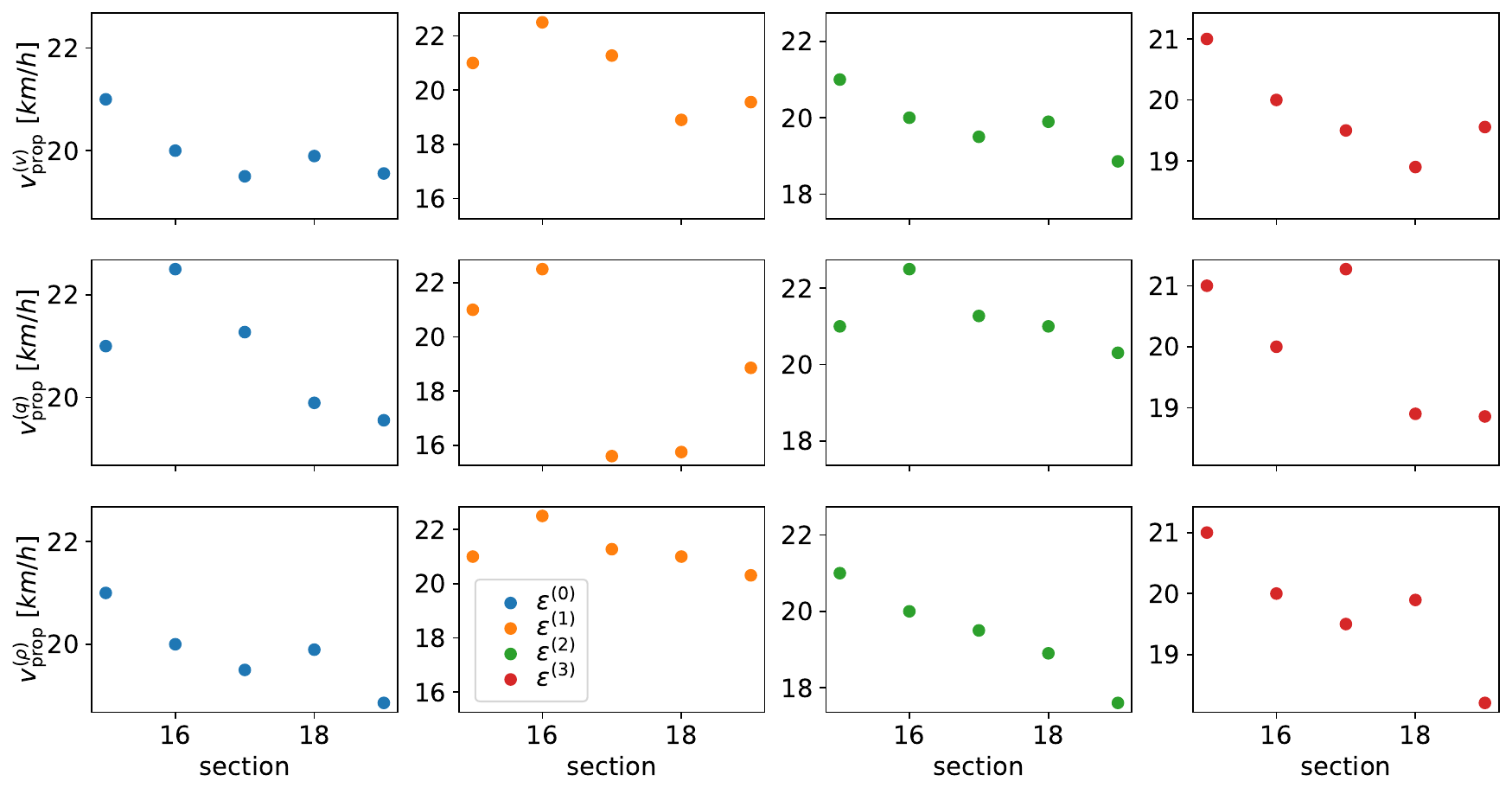}
		\caption{Motorway A3}
		\label{fig:prop_velo_A3}
	\end{subfigure}
	\caption{Propagation velocities $v_{\text{prop}}$ resulting from the minimal and maximal strength of the responses of each observable to each indicator at (a) interchange Breitscheid and (b) motorway A3. The original responses are depicted in \fig{fig:generic_shape_min}. The dashed line in (a) visually separates section 6 from the other sections due to its location.}
	\label{fig:prop_velo}
\end{figure}

Figure~\ref{fig:prop_velo} shows the propagation velocities for the two cases considered. Regarding the result of the Breitscheid interchange, we note that section 6 is a special case due to its location (see Secs.~\ref{sec3}~and~\ref{sec42}, and \fig{fig:Breitscheid_zoom} in Appendix~\ref{ap_map_breitscheid}), and is therefore separated by a dashed line in \fig{fig:prop_velo}~(a). In general, we find the values of the propagation velocities either very close to or within the typical interval of~$v_c$. In the majority of cases the values of $v_{\text{prop}}$ decrease into the numerical range of~$v_c$ with increasing distance to the indicator section~$j$. In other cases, the values of $v_{\text{prop}}$ derive from this behavior by scattering, e.g. $v_{\text{prop}}^{(\rho)}$ for Breitscheid in \fig{fig:prop_velo}(a) (last row) as well as $v_{\text{prop}}^{(q)}$ for motorway A3 in \fig{fig:prop_velo}(b) (second row). A comparison of these cases with the corresponding responses in \fig{fig:generic_shape} (or \fig{fig:generic_shape_min}) shows that the scattering can be ascribed to the characteristics of the responses. The shape and position of the local minima are not as clearly manifested as in other cases, making it difficult to accurately calculate the associated propagation velocities. Nevertheless, the numerical values are consistent with the typical range of~$v_c$. Unfortunately, it is not possible to derive a generic behavior on the basis of the results in \fig{fig:prop_velo}.

Considering that we derive the propagation velocities from the average response functions in a first approach, the results are meaningful. Although the velocities were determined separately for all indicators, the results are close to each other. While these results are not conclusive enough to derive a general behaviour regarding the dependence of the distance on the indicator section~$j$, the results provide a proof of concept. In the case of the classification of the congestion as a wide moving jam, the results do not argue against it.

\subsection{Scaling of responses}
\label{sec45}
 \begin{figure}[!htbp]
 	\centering
 	\begin{subfigure}[b]{1\textwidth}
 		\centering
 		\includegraphics[width=\textwidth]{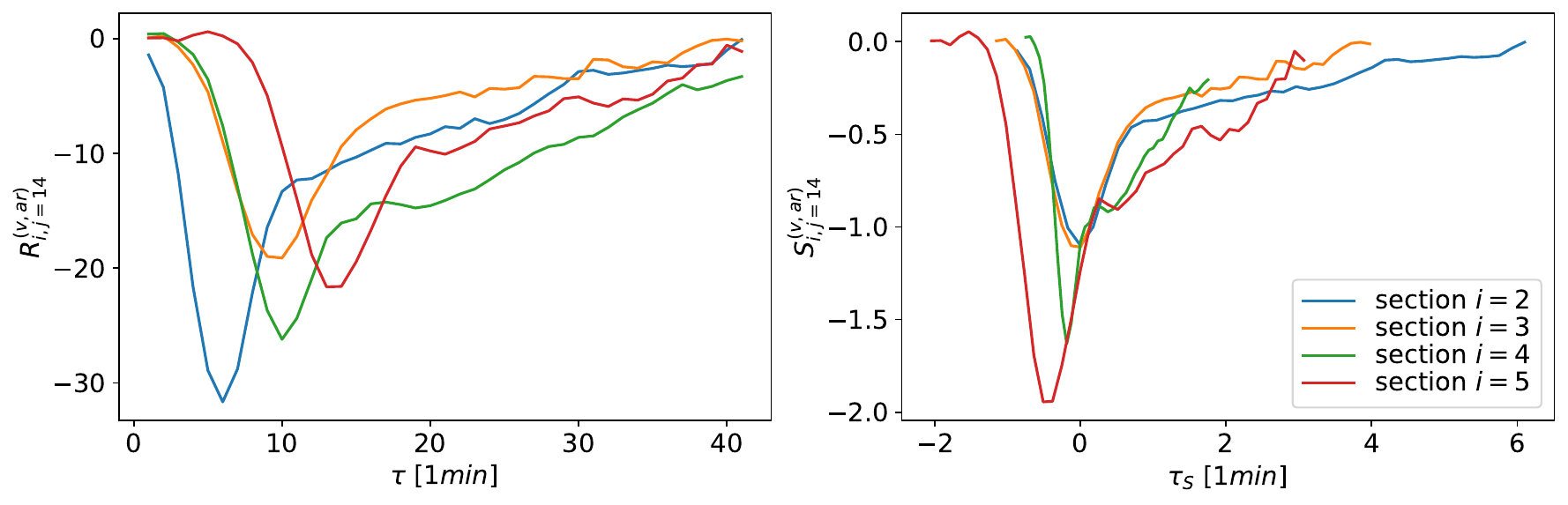}
 		\caption{Breitscheid}
 		\vspace{\baselineskip}
 		\label{fig:scaling_velo_A52}
 	\end{subfigure}
 	
 	\begin{subfigure}[b]{1\textwidth}
 		\centering
 		\includegraphics[width=\textwidth]{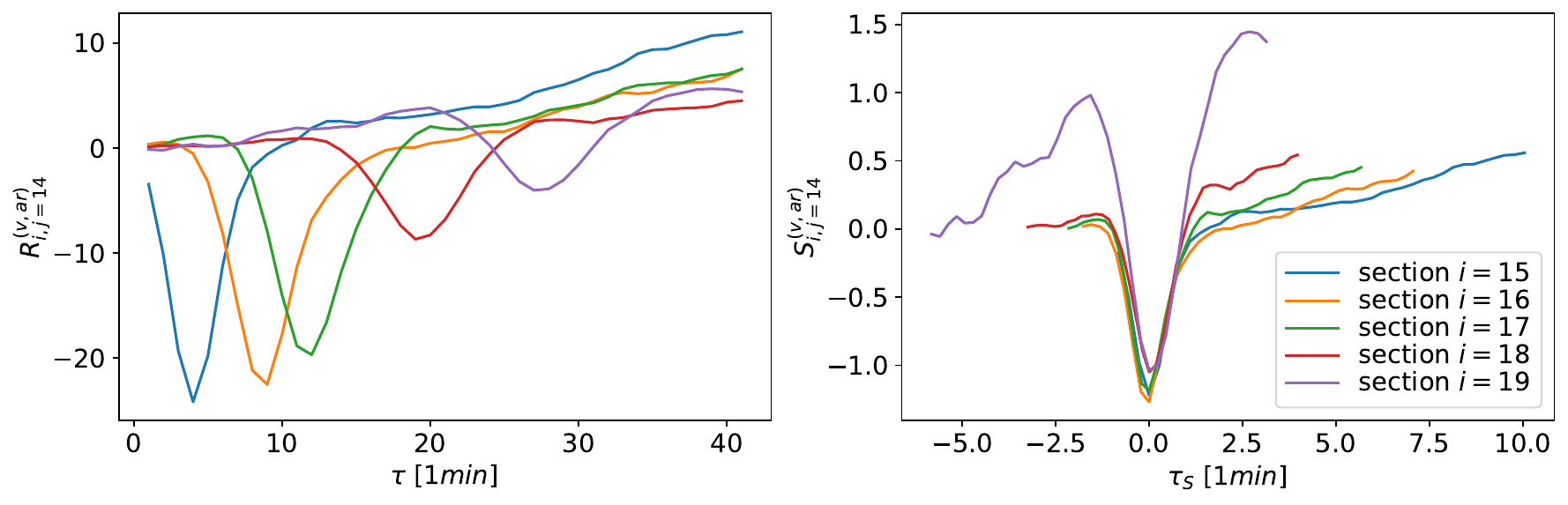}
 		\caption{Motorway A3}
 		\label{fig:scaling_velo_A3}
 	\end{subfigure}
 	\caption{Original (left column) and rescaled (right column) velocity responses of spatially subsequent sections~$i$ to congestion at (a) section $j=1$ at Breitscheid and (b) section $j=14$ on motorway A3 during afternoon rush hours for indicator $\indi{0}$.}
 	\label{fig:scaling_velo}
 \end{figure}

The similarities in shape of the responses for smaller time shifts are striking, prompting the question of scalability. We aim to determine a scaling behavior by applying the following approach. Let $\mu_i$ be the peak position of the i-th response curve and $\sigma_i$ the corresponding full width at half maximum. We define 

\begin{equation}
	\tau^{(i)}_s = \frac{\tau - \mu_i}{\sigma_i}
	\label{eq45_1}
\end{equation}

as a rescaled axis for the time shift $\tau$. In addition, we scale the responses with a factor $\alpha_i = 1/\sigma_i$ resulting in the scaling function

\begin{equation}
	S_i(\tau_s) = \frac{1}{\sigma_i} R_i(\tau^{(i)}_s).
	\label{eq45_2}
\end{equation}

Figure~\ref{fig:scaling_velo} provides examples of velocity responses and their respective rescaling in accordance with Eq.~\eqref{eq45_2} for the case of Breitscheid and motorway A3. The selected scaling method proves effective for both cases. Of course the success of the applied scaling method heavely depends on the manifestation of the minima, or maxima, respectively. In some cases it is not possible to properly scale our results. The determination of the full width half maximum $\sigma_i$ is difficult if the local extrema are not clearly shaped in the course of the responses. Nevertheless, the examples depicted in \fig{fig:scaling_velo} demonstrate a proof of concept and give a first glimpse at universality in the responses.

\section{Conclusion}
\label{sec5}

Traffic congestion affects millions of drivers daily, causing delays, frustration, and lost time. While this study focuses on the complex dynamics underlying congestion, the findings gained have direct practical value for every traffic participant. By quantifying how congestion propagates and how key traffic variables influence one another, we provide tools that can help improve traffic management, prediction, and control, ultimately aiming to reduce the time drivers spend stuck in traffic congestion.

Our analysis of response functions based on empirical data from heavily trafficked motorways in North Rhine-Westphalia revealed clear patterns of congestion propagation across several kilometers. The close interplay between velocity, flow, and density emerged naturally from the data, and response functions capture the oscillatory stop-and-go behavior familiar to many drivers. These response features showed distinct local minima and maxima within the first 30 minutes after congestion events, demonstrating the immediate impact and temporal evolution of traffic disturbances. By examining the width and shape of these features, we identified characteristic time scales that correspond to traffic oscillations and are consistent with phases described by three-phase traffic theory. This allowed us to differentiate congestion states and better understand their dynamics. Additionally, we derived estimates for the propagation velocity of congestion waves traveling through the motorway network. While these velocities align well with empirical observations, a universal relationship between propagation velocity and the distance from the initial congestion could not be identified. Another important finding is the discovery of a scaling property in the response functions, which causes a collapse of multiple response curves onto a single universal pattern. This hints at underlying regularities in traffic dynamics.

Several limitations influenced our study, primarily related to data quality, detector availability, and spatial constraints. The one-minute resolution of detector measurements limited the granularity of the analysis, and the spatial distribution of detectors restricted the ability to track congestion over larger areas. Moreover, averaging over time and lanes reduced the clarity of certain traffic patterns, making it challenging to precisely differentiate between congestion types or oscillation details. These factors affected the sharpness of local minima and maxima in the responses, which are critical for determining propagation velocities and scaling behavior. Despite these constraints, the response function approach offers a robust, data-driven alternative to more complex, parameter-heavy traffic models.

Importantly, the method holds significant promise for practical applications. Response functions provide a relatively simple yet powerful way to characterize local traffic dynamics and emerging patterns in real time. They reveal how traffic observables at specific road sections change based on prior congestion elsewhere, enabling better estimation, tracking, and potentially prediction of traffic conditions. Unlike traditional models, this approach does not require extensive calibration or assumptions about driver behavior, making it more adaptable and transparent. Further improvements, such as analyzing responses on a lane-by-lane basis or integrating more comprehensive detector networks, could increase accuracy and the ability to resolve detailed congestion phenomena. Such advances would directly support traffic engineers and planners in designing more effective traffic management strategies.

\section*{Acknowledgements}
We gratefully acknowledge the support from Deutsche Forschungsgemeinschaft (DFG) within the project “Korrelationen und deren Dynamik in Autobahnnetzen” (No. 418382724). We also thank Straßen.NRW for providing the empirical traffic data.

\section*{Author contributions}
M.S. and T.G. proposed the research. S.G. prepared the traffic data, performed all calculations, and wrote the manuscript with input from S.W., T.G. and M.S. All authors contributed equally to analyzing the results and reviewing the paper.

\section*{Declaration of Generative AI and AI-assisted technologies in the writing process}
During the preparation of this work the author(s) used ChatGPT in order to improve the language and readability. After using this tool/service, the author(s) reviewed and edited the content as needed and take(s) full responsibility for the content of the publication.

\bibliographystyle{unsrturl}

\addcontentsline{toc}{section}{References}
\clearpage

\begin{appendices}
\renewcommand{\thefigure}{A\arabic{figure}}
\setcounter{figure}{0}
\section{}
\subsection{Interchange Breitscheid: supplementary map}
\label{ap_map_breitscheid}
\begin{figure}[htbp]
	\begin{center}
		\includegraphics[width=0.9\textwidth]{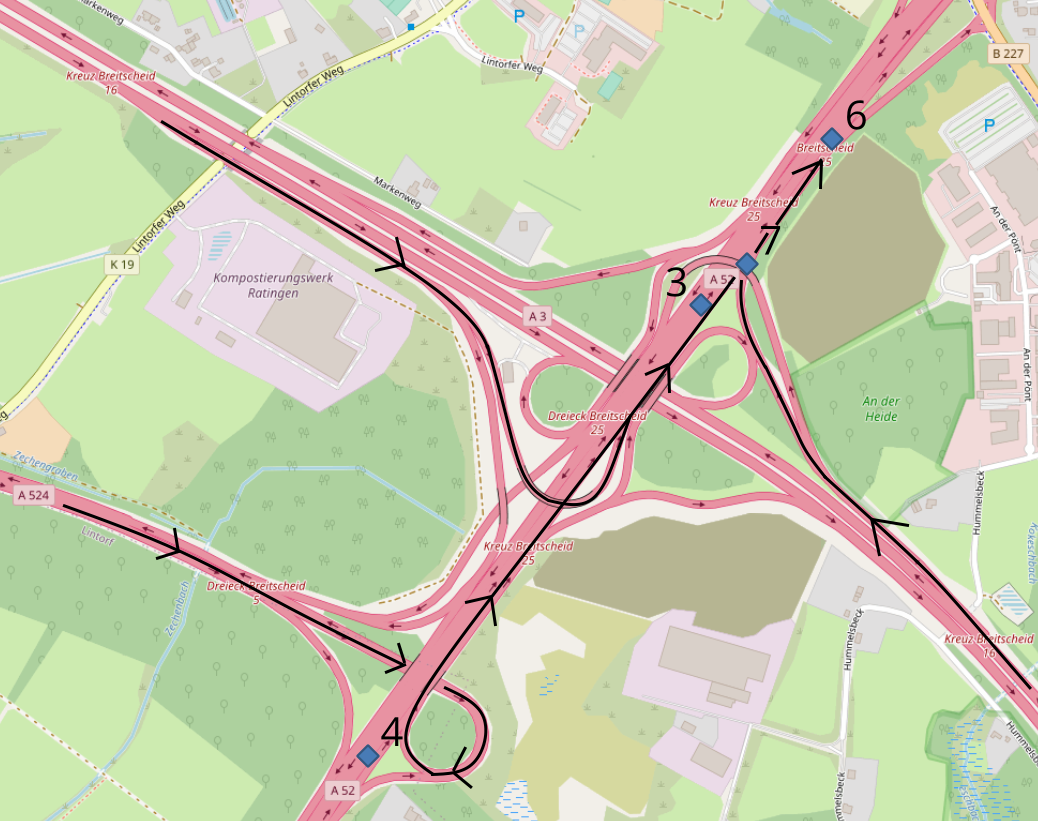}
		\caption{Zoom into the map of interchange Breitscheid shown in \fig{fig:map_locations}(a). Section 6 is located on a separated lane, merging traffic flows from connected motorways A524 and A3 indicated by black arrows.}
		\label{fig:Breitscheid_zoom}
	\end{center}
\end{figure}

Figure \ref{fig:Breitscheid_zoom} shows the details of the map for interchange Breitscheid in \fig{fig:map_locations}(a) to illustrate the special location of section 6 on a separate lane. The latter merges the incoming traffic flows from connected motorways A524 and A3 which is indicated by black arrows. After passing section 6, the traffic flow is merged into the lanes of motorway A52.

\clearpage

\subsection{Different representation of responses}
\label{ap_generic_shape_min}
\begin{figure}[htbp]
	\centering
	\begin{subfigure}[b]{0.94\textwidth}
			\centering
			\includegraphics[width=\textwidth]{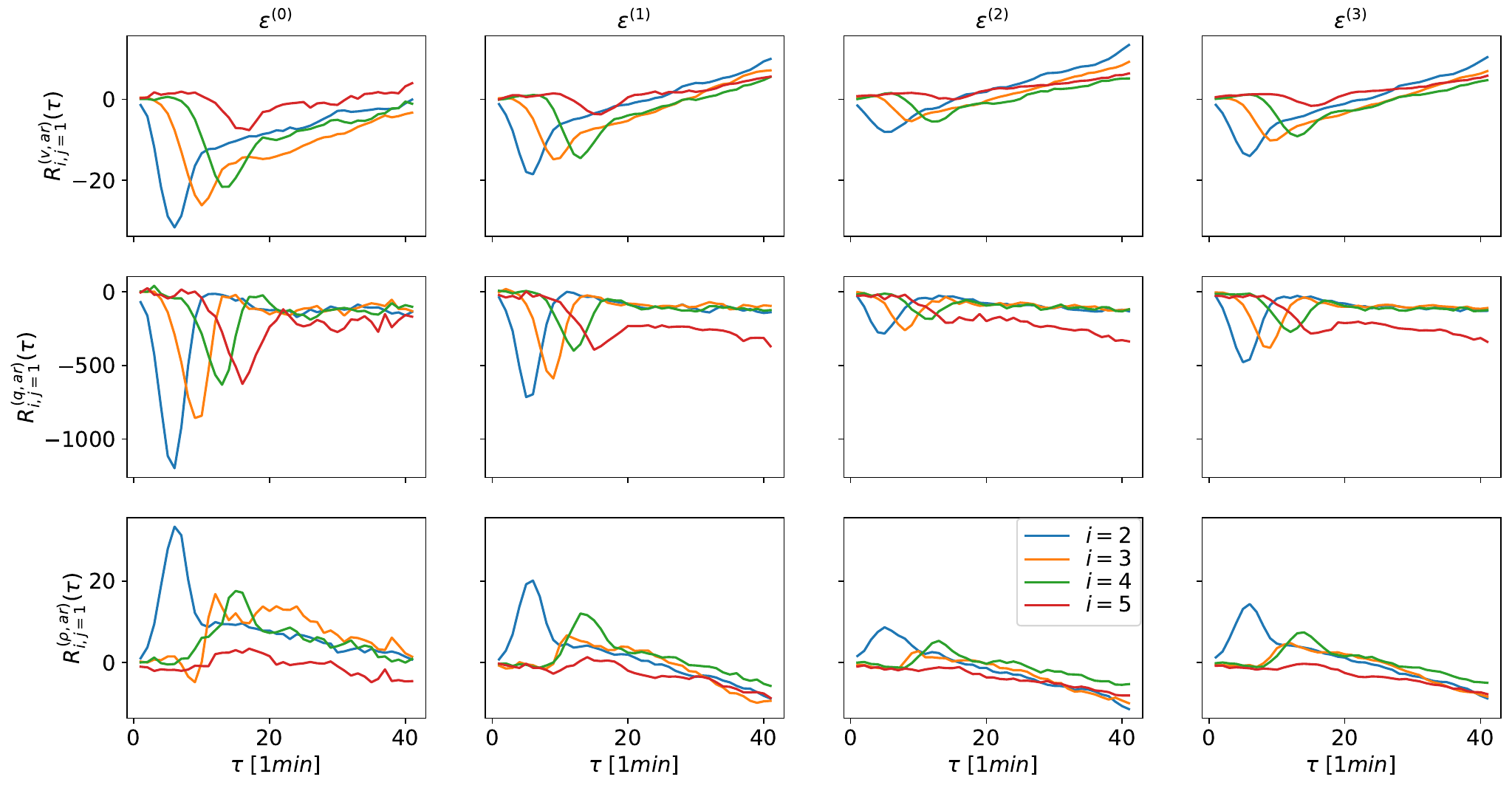}
			\caption{Breitscheid}
			\vspace{\baselineskip}
			\label{fig:generic_shape_min_A52}
		\end{subfigure}

	\begin{subfigure}[b]{0.94\textwidth}
			\centering
			\includegraphics[width=\textwidth]{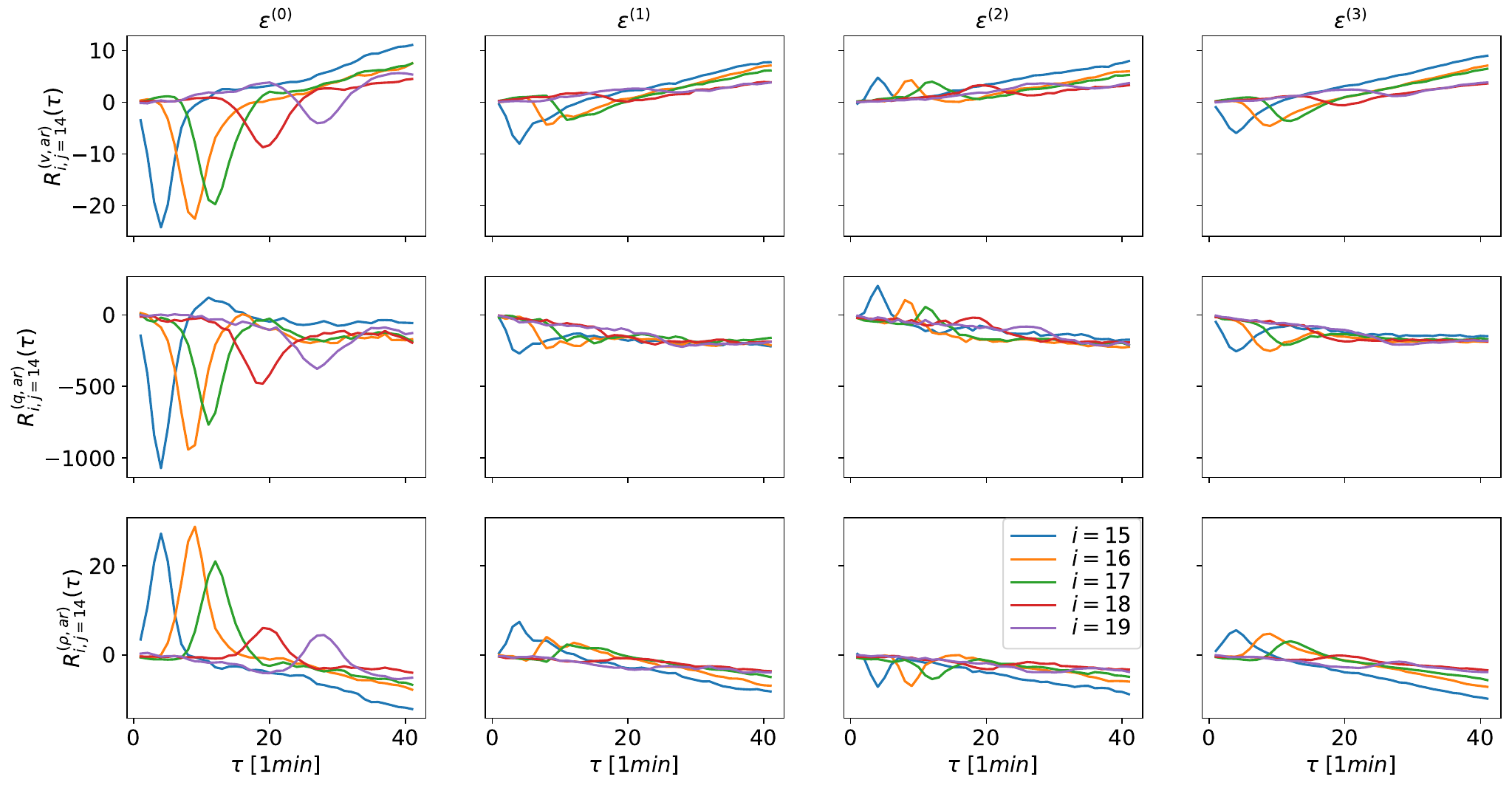}
			\caption{Motorway A3}
			\label{fig:generic_shape_min_A3}
		\end{subfigure}
	\caption{Velocity, flow and density response of spatially subsequent sections $i$ to congestion at (a) section $j=1$ at interchange Breitscheid during afternoon rush hours and (b) section $j=14$ on motorway A3 during afternoon rush hours broken down by the individual indicators $\varepsilon(t)$.}
	\label{fig:generic_shape_min}
\end{figure}

Figure \ref{fig:generic_shape_min} shows the velocity, flow and density responses of sections at interchange Breitscheid and motorway A3 broken down by individual indicators for congestion. It serves as supplemental information of \fig{fig:generic_shape} because it emphasizes the propagating nature of  the congestion in a more clear fashion.

\end{appendices}

\end{document}